# Niobium disulphide (NbS$_2$)-based (heterogeneous) electrocatalysts for an efficient hydrogen evolution reaction†

Leyla Najafi,‡[a] Sebastiano Bellani,‡[a] Reinier Oropesa-Nuñez, [b] Beatriz Martín-García, [a] Mirko Prato,[c] Vlastimil Mazánek,[d] Doriana Debellis,[e] Simone Lauciello,[e] Rosaria Brescia, [e] Zdeněk Sofer [d] and Francesco Bonaccorso [*ab]

The design of efficient and cost-effective catalysts for the hydrogen evolution reaction (HER) is the key for molecular hydrogen (H$_2$) production from electrochemical water splitting. Transition metal dichalcogenides (MX$_2$), most notably group-6 MX$_2$ (*e.g.*, MoS$_2$ and WS$_2$), are appealing catalysts for the HER alternative to the best, but highly expensive, Pt-group elements. However, their HER activity is typically restricted to their edge sites rather than their basal plane. Furthermore, their semiconducting properties hinder an efficient electron transfer to the catalytic sites, which impedes a high rate of H$_2$ production. Herein, we exploit liquid-phase exfoliation-produced metallic (1H, 2H and 3R) NbS$_2$ nanoflakes, belonging to the class of metallic layered group-5 MX$_2$, to overcome the abovementioned limitations. Both chemical treatment with hygroscopic Li salt and electrochemical *in operando* self-nanostructuring are exploited to improve the NbS$_2$ nanoflake HER activity. The combination of NbS$_2$ with other MX$_2$, in our case MoSe$_2$, also provides heterogeneous catalysts accelerating the HER kinetics of the individual counterparts. The designed NbS$_2$-based catalysts exhibit an overpotential at a cathodic current of 10 mA cm$^{-2}$ ($\eta_{10}$) as low as 0.10 and 0.22 V *vs.* RHE in 0.5 M H$_2$SO$_4$ and 1 M KOH, respectively. In 0.5 M H$_2$SO$_4$, the HER activity of the NbS$_2$-based catalysts is also superior to those of the Pt/C benchmark at current densities higher than 80 mA cm$^{-2}$. Our work provides general guidelines for a scalable and cost-effective exploitation of NbS$_2$, as well as the entire MX$_2$ portfolio, for attaining a viable H$_2$ production through electrochemical routes.

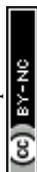

## Introduction

Molecular hydrogen (H$_2$) represents an ideal high-energy density source (between 120 and 140 MJ kg$^{-1}$),[1,2] since it can be produced *via* electrochemical water splitting from renewable sources[3,4] and its consumption is sustainable and environmentally friendly.[5,6] To speed up the rate of the hydrogen evolution reaction (HER) (4H$^+$ + 4e$^-$ → 2H$_2$ in acidic media; 4H$_2$O + 4e$^-$ → 2H$_2$ + 4OH$^-$ in alkaline media), commercial electrolyzers demand effective catalysts based on Pt-group elements.[7–9] However, the cost[10] and the scarcity[11] of these noble metal-based electrocatalysts hinder their massive use in a move toward a *hydrogen economy*.[12,13] Therefore, great efforts have been made towards the development of cost effective noble-metal-free alternatives, including transition metal alloys,[14,15] nitrides,[16,17] chalcogenides,[18–23] phosphides,[24,25] carbides,[26–28] and carbonaceous nanomaterials.[29–31] In this context, transition metal dichalcogenides –MX$_2$– (M = transition metal; X = S, Se, Te), composed of covalently bonded X–M–X blocks,[32,33] are appealing alternative catalysts for the HER.[34–38] Theoretical[39–41] and experimental studies[41–43] for the most investigated group-6 MX$_2$ (*e.g.*, H-MoS$_2$ and H-WS$_2$) have shown that the scarce unsaturated X-edges are the HER-active sites, since they have a near zero Gibbs free energy of adsorbed atomic H ($\Delta G_H^0$), while the more abundant basal plane is catalytically inert. Furthermore, the semiconducting properties of group-6 MX$_2$ limit the electron transfer to the catalytic edge sites,[44,45] thus hindering the high H$_2$ production rate required in electrolyzer systems. The design of nanostructured MX$_2$ (ref. 42 and 46–49) has been pursued to exploit their high *per site* activity, achieving overpotential *versus* the reversible hydrogen electrode (RHE) at 10 mA cm$^{-2}$ cathodic current density ($\eta_{10}$), approaching those of noble metal-based electrocatalysts (*i.e.*, <0.1 V).[28,50,51] However, MX$_2$ nanostructuring inevitably raises scalability and cost issues, impeding their facile implementation. Recently, metallic MX$_2$ based on group-5 metals (*i.e.*, vanadium (V), niobium (Nb), tantalum (Ta)) attracted utmost interest for the HER process due to the predicted HER activity of their basal planes (especially for sulphides),[52–58] beyond that of the chalcogen and metal edges.[55,56] Additionally, group-5 MX$_2$ have been theoretically proposed as ideal supports for anchoring single metal atom catalysts (*e.g.*, Pt, Ni and Pd), boosting the HER activity of the latter.[59] Recent experimental studies have validated the theoretical predictions, producing chemical vapour deposition (CVD)-synthetized 2H-TaS$_2$ and 2H-NbS$_2$ nanoplatelets displaying an $\eta_{10}$ of ~50–60 mV with a low catalyst loading of only 10–55 μg cm$^{-2}$.[54] Similar results have also been achieved by another NbS$_2$ polytype, *i.e.*, chemically exfoliated 3R-NbS$_2$ flakes.[57] However, the aforementioned performances were always reached after an electrochemical pre-conditioning of the electrodes consisting of thousands of cyclic voltammetry (CV) scans.[54,57] The latter cause a progressive self-optimizing evolution of the NbS$_2$ morphology, *i.e.*, a reduction in the thickness of the nanoplatelets.[54,57] This morphology change speeds up the electron transport towards the HER-active sites by shortening the interlayer electron-transfer pathways and facilitating the access of the aqueous protons (H$_3$O$^+$) to the catalytic films.[54,57] Meanwhile, it increases the electrochemically accessible surface area of the electrode films.[54,55,57] Parallel investigations also claimed that during the cycling process, the oxides naturally formed on the ambient-exposed H-TaS$_2$ surface are peeled off by the H$_2$ bubbles and the real HER activity of H-TaS$_2$ is therefore exhibited subsequently.[55] Despite these breakthrough performances, such self-optimizing fragmentation of the catalyst films could negatively affect their adhesion to the electrode substrates, raising severe doubts about their reliability for high-rate H$_2$ production.[60] Therefore, novel insight into the processing of metallic group-5 MX$_2$ is urgently required for their practical validation for the HER. Prospectively, optimized electrode morphologies obtained during the deposition of the electrode films could be "freezed" by using catalyst binders, such as sulfonated tetrafluoroethylene-based fluoropolymer copolymers (*e.g.*, Nafion).

[a]*Graphene Labs, Istituto Italiano di Tecnologia, Via Morego 30, 16163 Genova, Italy. E-mail: francesco.bonaccorso@iit.it*

[b]*BeDimensional Spa., Via Albisola 121, 16163 Genova, Italy*

[c]*Materials Characterization Facility, Istituto Italiano di Tecnologia, Via Morego 30, 16163 Genova, Italy*

[d]*Department of Inorganic Chemistry, University of Chemistry and Technology Prague, Technická 5, 166 28 Prague 6, Czech Republic*

[e]*Electron Microscopy Facility, Istituto Italiano di Tecnologia, Via Morego 30, 16163 Genova, Italy*

† Electronic supplementary information (ESI) available. See DOI: 10.1039/c9ta07210a

‡ These authors contributed equally.

In this work, we report the HER activity of single/few-layer NbS$_2$ flakes produced by an environmentally friendly liquid phase exfoliation (LPE) of synthetized material crystals (mixture of 3R- and 2H-polytypes). In contrast to CVD growth[61] and mechanical exfoliation,[62] LPE is promptly scalable[63–68] and does not require expensive nanostructuring of NbS$_2$ materials. Moreover, LPE does not use hazardous chemicals,[69] which could affect the physico-chemical properties of the chemically exfoliated materials.[70–72] Finally, LPE does not face the issues related to electrochemical exfoliation methods, *e.g.*, the material oxidation during anodic exfoliation in aqueous electrolytes and the ion intercalation-induced morphology degradation effects.[73,74] To improve the catalytic activity of the NbS$_2$ flakes, a solution-based chemical treatment with hygroscopic bis(trifluoromethane)sulfonimide lithium salt (Li-TFSI) was used to enhance the electrolyte accessibility to the HER-active surface of NbS$_2$ flakes, overcoming the electrically insulating oxidation of the flake surfaces.[75,76] *In operando* electrochemical cycling treatment, as recently reported in the literature[54,55] and licensed document,[77] was also applied to the LPE-produced NbS$_2$ flakes. Furthermore, the chemical and the electrochemical treatments were combined to positively benefit from possible synergistic effects on the HER activity of the electrodes. Finally, on the basis of published density functional theory (DFT) calculations and *ab initio* molecular dynamics (AIMD) simulations,[56] NbS$_2$ flakes were hybridized with MoSe$_2$ flakes to tune the $\Delta G_H^0$ of the basal planes of NbS$_2$ flakes to optimal thermo-neutral values (*i.e.*, 0 eV) in the resulting heterogeneous configurations for both edge and basal sites. In acidic solution (0.5 M H$_2$SO$_4$), the designed heterogeneous group-5 TMD catalysts (named NbS$_2$:MoSe$_2$) indeed outperform their individual components ($\eta_{10}$ of 0.10 V for NbS$_2$:MoSe$_2$, 0.42 V for NbS$_2$ and 0.28 V for MoSe$_2$). For the first time, the HER activity was also investigated in alkaline (1 M KOH) solutions, showing a promising $\eta_{10}$ of 0.22 V. The HER activity of our best electrocatalysts was tested over 12 h of continuous operation at a fixed potential corresponding to an initial cathodic current density of 80 mA cm$^{-2}$, validating the electrochemical durability of NbS$_2$-based electrocatalysts. Our results aim to provide key guidelines to efficiently exploit two-dimensional (2D) NbS$_2$ and, more generally, 2D metallic group-5 MX$_2$ for the HER *via* scalable material and electrode processing.

## Results and discussion

### Synthesis and exfoliation of NbS$_2$ crystals

NbS$_2$ crystals were synthesized by direct reaction from Nb and S elements.[78] In more detail, Nb powder and S granules at a *ca.* 1 : 2 elemental stoichiometry (1 wt% excess of S to avoid Nb-intercalated structures,[79]) were loaded in a quartz glass ampoule were heated at 450 °C for 12 h and subsequently at 600 °C for 48 h. The products were then treated at 900 °C for 48 h and cooled down to room temperature over a period of 24 h. The as-produced NbS$_2$ crystals were characterized by scanning electron microscopy (SEM) coupled with energy-dispersive X-ray spectroscopy (EDS) (Fig. 1a–c), revealing a near-ideal stoichiometric composition of the NbS$_2$ crystals (S : Nb atomic ratio ~1.8, see ESI, Table S1†), which is in agreement with previous studies.[78] The layered structure of the NbS$_2$ crystals is clearly visible on the edges, as shown by the SEM image in Fig. 1d. The NbS$_2$ flakes were produced by LPE[63,80] of the synthetized crystals in 2-propanol (IPA) followed by sedimentation-based separation (SBS)[81,82] to remove un-exfoliated crystals by ultracentrifugation (see the Experimental section). The use of IPA as a solvent is effective to remove possible S impurities from the sample during the LPE, SBS and filtration processes in the supernatant, since S is slightly soluble in alcohols (solubility > 0.05 mg mL$^{-1}$).[83] Moreover, starting from cost-effective artificial crystals, the LPE method does not require complex material processing or time-consuming bottom-up nanomaterial synthesis (*e.g.*, CVD).[84] The morphology of the as-produced NbS$_2$ flakes was characterized

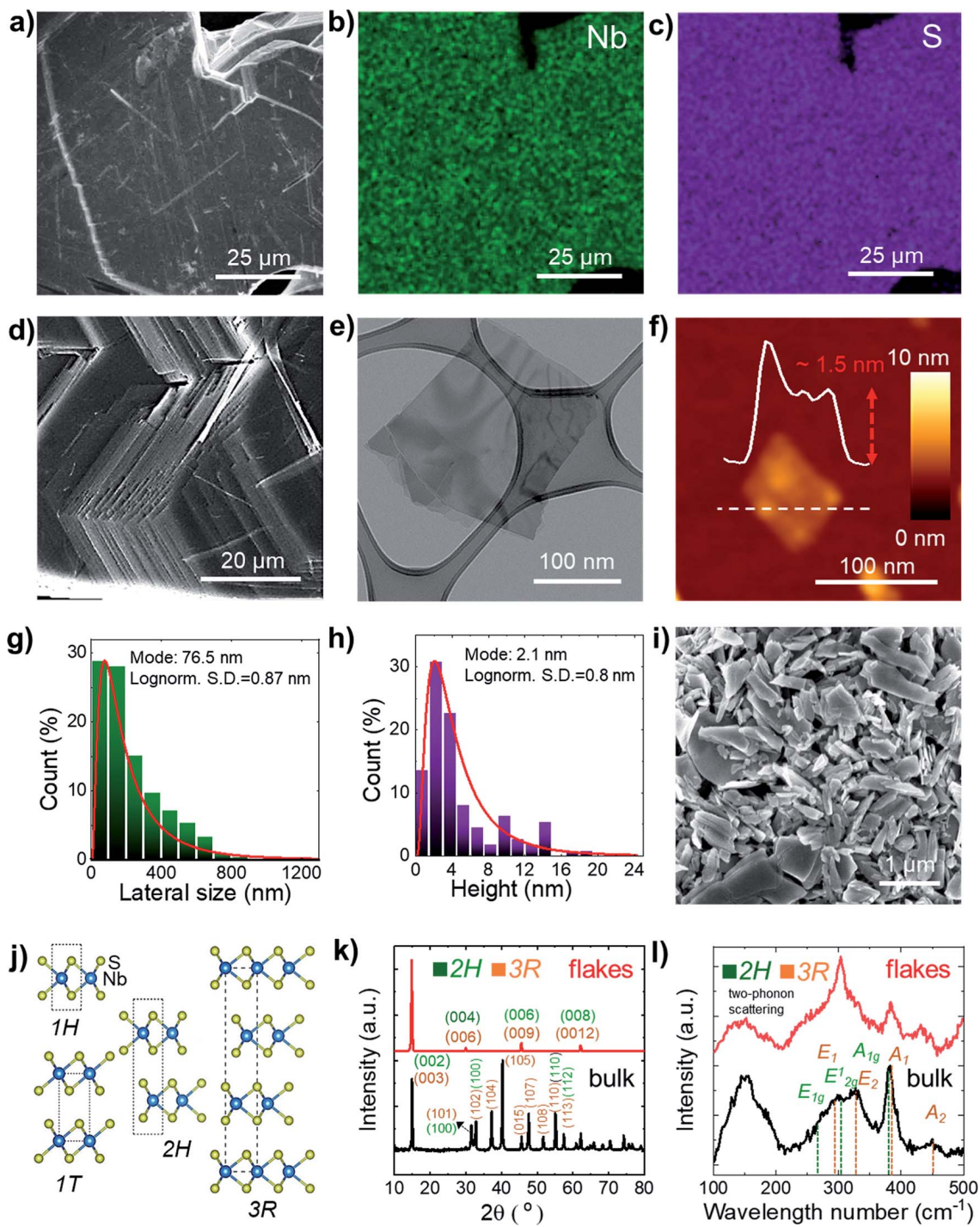

**Fig. 1** (a) SEM image of the as-synthetized NbS$_2$ crystals and the corresponding quantitative EDS maps for (b) Nb (L$\alpha$ peak at 2.17 keV, in green) and (c) S (K$\alpha$ peak at 2.31 keV, in violet). (d) SEM image of a NbS$_2$ crystal edge showing its layered structure. (e) BF-TEM image of LPE-produced NbS$_2$ flakes. (f) AFM image of LPE-produced NbS$_2$ flakes. (g) BF-TEM-based statistical analysis of the lateral dimension of the NbS$_2$ flakes. (h) AFM statistical analysis of the NbS$_2$ flakes. (i) SEM image of a film of NbS$_2$ flakes filtered through a microporous nylon membrane, showing the presence of NbS$_2$ flakes with a lateral size >1 μm. (j) Structure of the NbS$_2$ polytypes experimentally observed in the literature.[79] (k) XRD and (l) Raman spectra of the synthesized NbS$_2$ crystals and the LPE-produced NbS$_2$ flakes. The XRD and Raman peaks assigned to 2H- and 3R-NbS$_2$ are also shown.

by bright-field transmission electron microscopy (BF-TEM) and atomic force microscopy (AFM) in order to evaluate their lateral dimension and thickness, respectively. Fig. 1e shows the BF-TEM image of representative NbS$_2$ flakes, displaying wrinkled surfaces with irregular shapes, but sharp edges. Fig. 1f shows an AFM image of a representative NbS$_2$ flake, together with its height profile showing a step edge of ~1.5 nm.

Statistical BF-TEM-based analysis (Fig. 1g) indicates that the lateral size of the flakes follows a log-normal distribution peaked at ∼76 nm, with maximum values above 1 μm. The statistical AFM analysis (Fig. 1h) shows that the sample is mainly composed of few-layer $NbS_2$ flakes (the AFM thickness of a $NbS_2$ monolayer lies generally between 0.6 nm and 0.8 nm, depending on the substrate and the AFM instrumentation;[85–87] the $NbS_2$ interlayer distance is ∼0.6 nm (ref. 88–91)), with a log-normal distribution peaked at ∼2.1 nm.

Additional SEM measurements were carried out on $NbS_2$ flakes filtered through a microporous nylon membrane (Fig. 1i). Such a material deposition method produce films exhibiting $NbS_2$ flakes with a lateral size larger than 1 μm. The crystal structure of the $NbS_2$ crystals and flakes was evaluated by X-ray diffraction measurements (XRD). Fig. 1j shows the structure of the $NbS_2$ polytypes experimentally observed in the literature.[79] Typically, $NbS_2$ crystals are found in two possible hexagonal polytypes with a distinctive stacking order of the $NbS_6$ prisms.[92,93] The first one, 2H-$NbS_2$, is composed of two $NbS_2$ layer per unit cell (space group: $P6_3/mmc$), whereas the second one, 3R-$NbS_2$, consists of three layers per unit cell (space group: $R3m$).[92] Other phases of $NbS_2$ are thermodynamically unfavoured (trigonal 1T-$NbS_2$; space group: $P\bar{3}m1$, derived under special deposition conditions in a monolayer or thin film form),[94,95] or only theoretically predicted (Haeckelite 1S-$NbS_2$; space group: $P4/mbm$;[96] at pressure >20 GPa, and tetragonal $NbS_2$; space group: $I4/mmm$ (ref. 93)). The XRD data (Fig. 1k) show that the as-synthetized $NbS_2$ crystals consist of a mixture of both 2H and 3R polytypes (indexed to PDF cards no. 04-005-8447 (ref. 97 and 98) and no. 04-004-7343,[91,98] respectively). After exfoliation, the sample shows XRD peaks corresponding to the diffractions on $NbS_2$ crystals aligned along the (001) plane[99,100] (or to 1H-$NbS_2$, *i.e.*, the monolayer form of both 2H and 3R-$NbS_2$). This indicates that the $NbS_2$ flakes are arranged parallel to the c-axis perpendicular to the substrate and retain their native crystal structure.[48,101] Raman spectroscopy analysis (Fig. 1l) further confirms the crystallinity of the exfoliated samples, well displaying the nondegenerate Raman active modes of the native crystals predicted by group theory for 2H-$NbS_2$ (space group: $D_{6h}^4$)[102,103] and 3R-$NbS_2$ (space group: $C_{3v}^5$).[102,104] In more detail, 2H-$NbS_2$ exhibits the peaks attributed to $E_{1g}$, $E_{2g}^2$ and $A_{1g}$ modes at ∼265, ∼305 cm$^{-1}$ and ∼380, respectively,[102,103] while 3R-$NbS_2$ shows the ones associated to $E_1$, $E_2$, $A_1$ and $A_2$, at ∼290, ∼330, ∼385, and ∼450 cm$^{-1}$, respectively.[102,104–106] The broad peaks observed at ∼160 and ∼180 cm$^{-1}$ originate from two-phonon scattering processes in the presence of defects.[86,91,102,104] Notably, $NbS_2$ flakes show a prevalence of the characteristic Raman peaks assigned to 2H-$NbS_2$. This suggests that the exfoliation of 3R-$NbS_2$ crystals intrinsically causes a 3R-to-H phase conversion toward bilayer and monolayer flakes (*i.e.*, 2H-$NbS_2$ and 1H-$NbS_2$), in agreement with previous observations.[86] In addition, the $A_2$ mode of 3R-$NbS_2$ redshifts from ∼450 cm$^{-1}$ in the $NbS_2$ crystal to ∼435 cm$^{-1}$ in the $NbS_2$ flakes, as a consequence of the relaxation of the interlayer van der Waals forces with decreasing the $NbS_2$ thickness.[86,107] The prevalence of the 2H-phase in the $NbS_2$ flakes was further confirmed by high resolution TEM (HRTEM) analysis. Fig. 2a shows a BF-TEM image of a partially suspended $NbS_2$ flake. The corresponding HRTEM image of a portion of the suspended region is shown in Fig. 2b, which shows a single-crystal structure. Fig. 2c shows the corresponding fast Fourier transform (FFT), which matches with that of [001]-oriented 2H-$NbS_2$ (ICSD card no. 250595 (ref. 108)). Fig. 2d shows a magnified portion of the HRTEM image, indicating lattice planes. Scanning transmission electron microscopy coupled with EDS (STEM-EDS) analyses were carried out to evaluate the chemical quality of the $NbS_2$ flakes. Fig. 2e shows the high-angle annular dark-field STEM (HAADF-STEM) image of a partially suspended $NbS_2$ flake. Fig. 2f and g show the corresponding STEM-EDS maps of Nb and S. The quantitative elemental analysis results in a S/Nb atomic ratio of ∼1.8 and a low atomic content of O (O/Nb atomic ratio ∼0.1), which excludes a predominant presence of oxide domains on the flakes. X-ray photoelectron spectroscopy (XPS) measurements (Fig. S1†) were performed to further analyse the element oxidation states on the surface of both the $NbS_2$ crystal and flakes, revealing the presence multiple $NbS_2$ phases, *i.e.*, 2H (or 1H) and 3R ones. In particular, the Nb 3d spectra show the presence of three doublets. The first doublet (peaks at 203.4 ± 0.2 eV and 206.1 ± 0.2 eV) and the second doublet (peaks at 204.0 ± 0.2 and 206.7 ± 0.2 eV) may be both assigned to $Nb^{4+}$ states in $NbS_2$.[107,109–111] The origin of two doublets for $NbS_2$ could be ascribed to the presence of multiple $NbS_2$ phases, *i.e.*, 2H (or 1H) and 3R ones. The peaks located at binding energies of 207.7 ± 0.2 eV and 210.4 ± 0.2 eV are assigned to the Nb(5+) state in $Nb_2O_5$.[112–116] Both the 3R phase and oxidization are attributed to the surface or the edges of the flakes because they were not detected using HRTEM and STEM-EDS (oxygen map not shown here, due to the low oxygen content), which are less surface-sensitive than XPS.

Overall, the morphological, structural and elemental analyses discussed above have shown that $NbS_2$ flakes were successfully produced by LPE in IPA starting from synthetized $NbS_2$ crystals. The $NbS_2$ flakes optimally retained the crystallinity of the native crystal. However, their surface oxidation has to be considered for exploring effectively their electrocatalytic properties for the HER,[78] in agreement with theoretical expectations.[52–56]

## Fabrication and electrochemical characterization of $NbS_2$-based electrodes

The direct exploitation of $NbS_2$ as an electrocatalytic material for the HER has been reported only recently exclusively under acidic conditions.[58,78,109,117,118] However, just two studies on 2H-$NbS_2$ nanoplatelets[54] and mechanically exfoliated tens of nm-thick 3R-$NbS_2$ fakes[57] have shown the possibility to approach Pt-like performance after thousands of electrochemical CV cycles, in agreement with theoretical predictions.[52–57] The unusual self-optimizing/nanostructuring behaviour of $NbS_2$ nanoplatelets inevitably raises questions about their reliability for a long-term HER (see the discussion in the Introduction section), pointing out the importance to provide new understanding for the fully exploitation of group-5 $MX_2$.[60]

The electrodes were advantageously produced through a room temperature vacuum filtration of the as-produced $NbS_2$

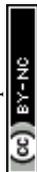

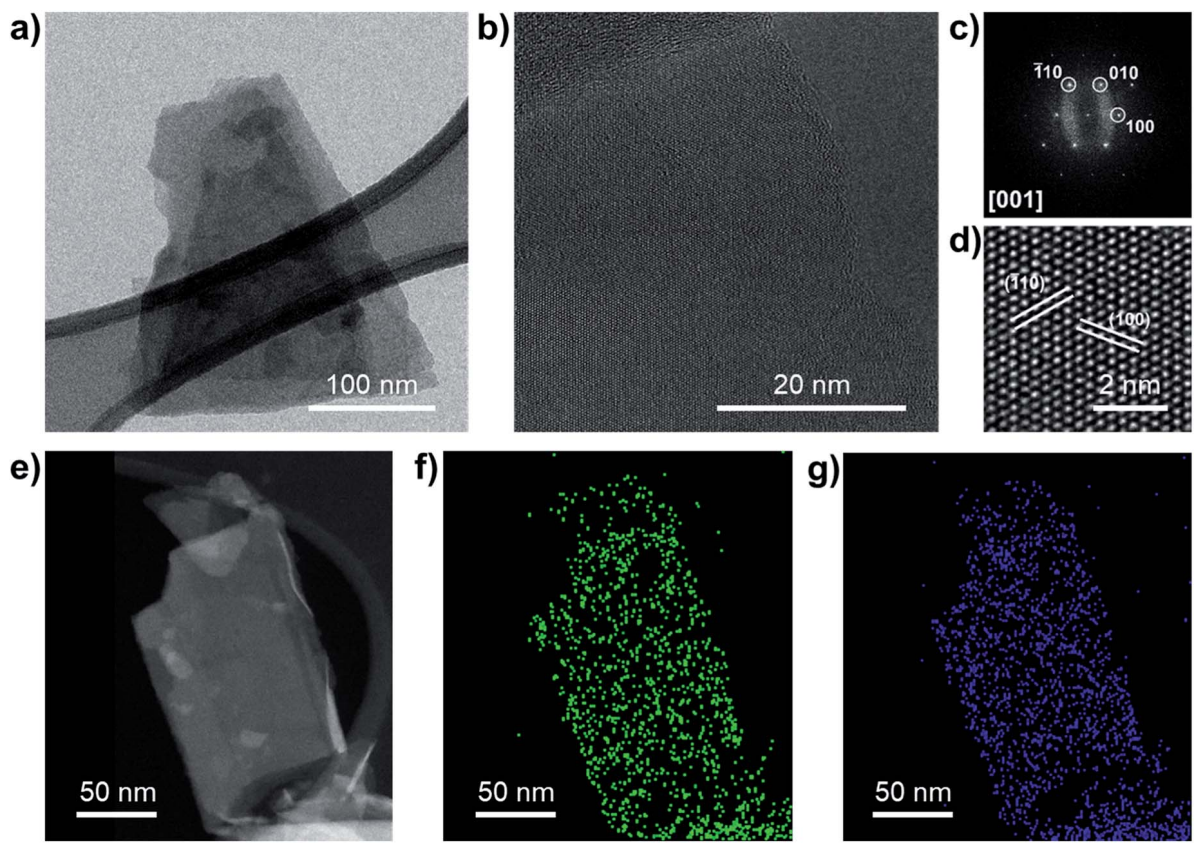

Fig. 2 (a) Elastically filtered BF-TEM image of a NbS$_2$ flake, partially suspended on a hole in the carbon support film. (b) HRTEM image of a portion of the suspended region of the NbS$_2$ flake, with (c) the corresponding FFT matching with that of [001]-oriented 2H-NbS$_2$. (d) Magnified portion of the HRTEM image. The lattice planes are also indicated. (e) HAADF–STEM image of a partially suspended NbS$_2$ flake and (f and g) corresponding quantitative STEM-EDS maps of Nb (Kα, in green) and S (Kα, in blue), respectively.

flake dispersions (catalyst mass loading, ∼0.20 mg cm$^{-2}$) on top of a single-walled carbon nanotube (SWCNT) buckypaper (SWCNT mass loading, ∼ 1.31 mg cm$^{-2}$). In fact, by taking advantage of the material topologies, the NbS$_2$ flakes cannot pass through the SWCNT tangle.[119] Consequently, the SWCNTs effectively filter the NbS$_2$ flakes, which then form a homogeneous film onto the buckypaper. The preparation of the SWCNT dispersion and the details of the fabrication of the electrodes are described in the Experimental section, following similar protocols adopted in our previous studies on LPE-produced transition metal (di or mono)chalcogenide-based electrocatalysts.[21,28,48,50,51] Fig. 3a shows a photograph of a representative NbS$_2$ electrode, which was manually bent to show its flexibility. The surface morphology of the as-prepared electrodes was characterized by SEM measurements. The NbS$_2$ flakes uniformly cover the mesoporous network of SWCNTs (Fig. 3b), which form a bundle-like morphology (Fig. S2†). The top-view SEM enlargement (Fig. 3c) shows a similar electrode surface to that displayed by NbS$_2$ flakes directly filtered through a nylon membrane without SWCNTs (see Fig. 1i). To improve the catalytic properties of the NbS$_2$ electrodes, a solution-based chemical treatment with hygroscopic Li-TFSI salt was used to enhance the electrolyte accessibility to the HER-active surface of the NbS$_2$ flakes (Fig. 3d). In more detail, the salt ions penetrate into the porous electrodes, occupying *inter*-flakes sites. The Li$^+$ can also intercalate into both surface niobium oxides (*i.e.*, Nb$_2$O$_5$ and NbO)[120–123] and NbS$_2$,[98,124–126] thus filling material interstitial sites. Concurrently, the Li-TFSI salt is highly hygroscopic, which means that it strongly adsorbs water, as shown in singular aqueous lithium-ion chemistry.[127] The water uptake was confirmed by four-probe resistivity measurement of films of NbS$_2$ flakes, showing a sheet resistance of ∼5.1 × 10$^2$ kΩ □$^{-1}$ after Li-TFSI treatment, whereas the untreated films exhibited a sheet resistance higher than 10$^3$ kΩ □$^{-1}$ due to the surface oxidation-induced passivation of the *inter*-flake electrical contact. During the air ambient exposure of the NbS$_2$-based electrodes, Li-TFSI can also diffuse outward, as also observed in Li-TFSI-doped organic semiconducting films commonly exploited in perovskite solar cells.[128,129] In addition, Li$_2$O or LiOH can also be formed by exposing Li-TFSI-treated electrodes to air or water,[130,131] resulting in H$_2$ gas evolution. Therefore, the water dragging, the outward diffusion of Li-TFSI and the H$_2$ evolution may drive the reorientation and nanostructuring (*i.e.*, exfoliation and lateral dimension reduction) of the NbS$_2$ flakes (Fig. S3†). These effects facilitate the water accessibility to the catalytic surface of the NbS$_2$ flakes. Notably, the effect of the Li-TFSI treatment resembles the H$_2$ evolution-aided nanostructuring of the group-5 MX$_2$-based electrodes occurring

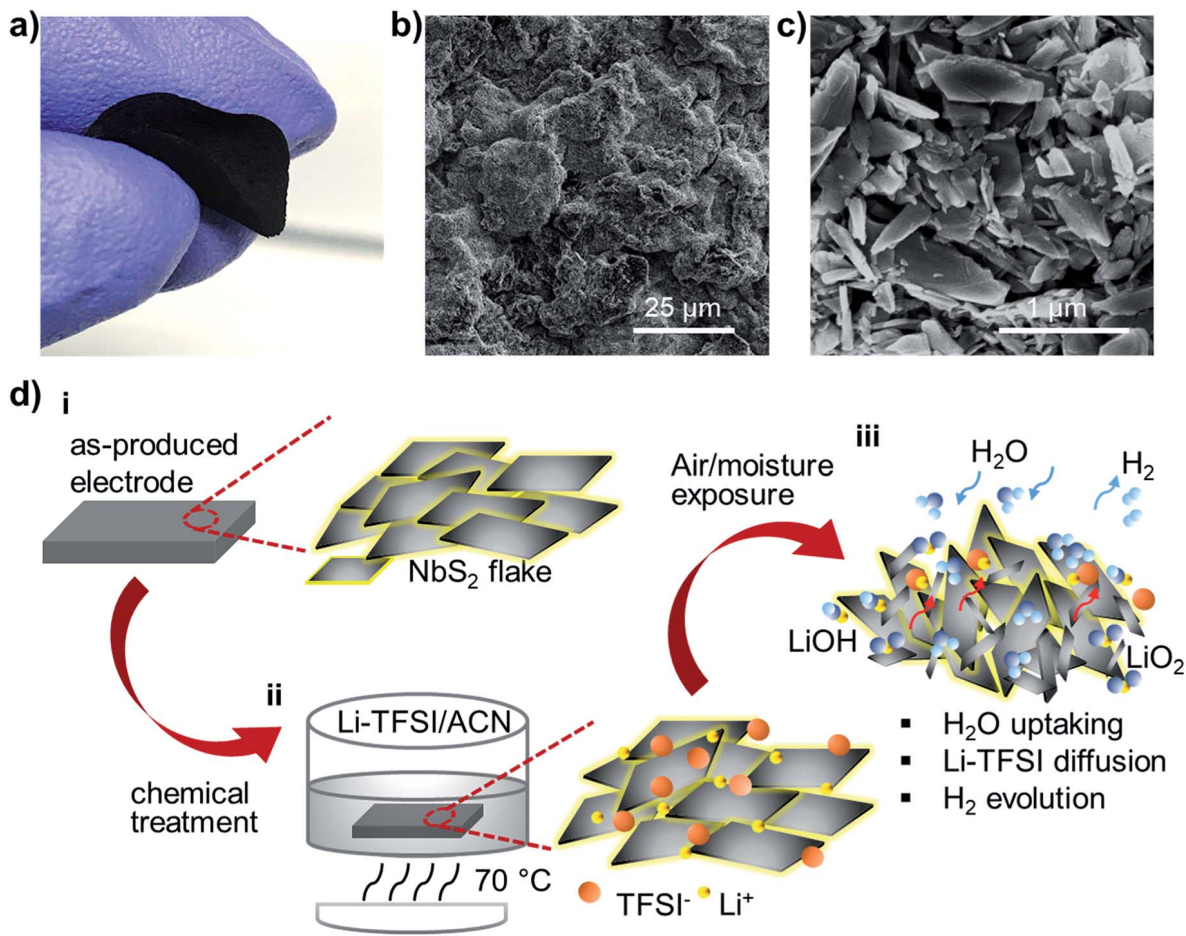

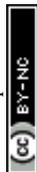

**Fig. 3** (a) A photograph of a flexible self-standing $NbS_2$-based electrode produced by room-temperature sequential vacuum filtration of SWCNT and $NbS_2$ flake dispersions onto a nylon membrane. (b and c) Top-view SEM images of a $NbS_2$-based electrode. (d) Sketch of the chemical treatment of the $NbS_2$-based electrodes: the as-produced $NbS_2$-based electrode (panel i) and that immersed in a Li-TFSI solution in ACN and heated at 70 °C (panel ii). The salt ions ($Li^+$ and $TFSI^-$) penetrate between the flakes and/or intercalate into surface oxide or $NbS_2$, forming a hygroscopic electrode surface. The latter takes up the water, Li-TFSI diffuses towards the surface, and $Li^+$ species react with air and water to form $Li_2O$ and LiOH and evolve $H_2$ (panel iii), reorienting the $NbS_2$ flakes to increase the water accessibility to the electrode.

during electrochemical treatments (*e.g.*, CV scans).[54,55] As reported in previous studies,[54,55,57] the $H_2$ evolution during CV scans causes a progressive self-optimizing evolution of the $NbS_2$ morphology from thick-to-thin flakes. This change of the $NbS_2$ morphology speeds up the electron transport towards the HER-active sites by shortening the interlayer electron-transfer pathways and facilitating the access of the aqueous protons ($H_3O^+$) to the catalytic films.[54,55,57] Meanwhile, it increases the electrochemically accessible surface area of the electrode films.[54,55,57] However, in contrast to our chemical approach, the electrochemical preconditioning of the electrodes is time-consuming and does not allow any reliable control of the electrode morphology from the start of the HER operation.

The modification of the catalyst film morphology (deposited on flat glassy carbon substrates) induced by both chemical and electrochemical treatments was verified by measuring the double layer capacitances ($C_{dl}$) of the films (Fig. S4a†). The $C_{dl}$ is proportional to the electrochemically accessible surface area of the films. These results show that the $C_{dl}$ of the Li-TFSI-treated catalyst film ($\sim$10.8 mF cm$^{-2}$) is more than three orders of magnitude higher than that of the untreated film. After the electrochemical treatment (*i.e.*, 1000 CV scans, see details in the Experimental section) in 0.5 M $H_2SO_4$, the catalyst film lost a significant amount of material, which was visible to the naked eye (Fig. S4b†). This effect could be a consequence of the catalyst film fragmentation originated from the $H_2$ bubbling,[54,55,57] as well as by the dissolution of possible surface oxides, which can form $Nb(OH)_4^-$ in acidic media.[132] However, the $C_{dl}$ of the electrochemically treated $NbS_2$ film is similar to that of the untreated film, indicating that the CV scans increase the specific electrochemically accessible surface area (defined by the ratio of the electrochemically accessible surface area and the mass loading) of the as-produced electrodes. X-ray photoelectron spectroscopy measurements (Fig. S5†) do not reveal significant changes in the chemical composition of the $NbS_2$ flakes after both the chemical and the electrochemical treatments, indicating that the latter mainly affect the morphology of the catalysts. Beyond the aforementioned morphology changes, it is worth noticing that the presence of alkali-metal cations, including $Li^+$, can control the catalytic activity of a catalyst in alkaline media.[133–135] In fact, it has been suggested that $Li^+$ induces steric and/or electronic effects on the interfacial water structure and reactivity, enhancing the HER activity of catalysts anchoring $Li^+$ in the form of complexes.[133] Noteworthily, the effective interaction between the Li-TFSI-treated electrode and the electrolyte is also supported by $C_{dl}$ measurements (Fig. S4†), which show an efficient electrochemical double layer formation (Fig. S5†).

### Electrochemical measurements of the electrodes

The HER activity of the as-produced electrodes was evaluated in both acidic (0.5 M $H_2SO_4$) and alkaline (1 M KOH) $N_2$-purged solutions at room temperature. To the best of our knowledge, no previous study investigated the HER activity of $NbS_2$ in alkaline solutions, either theoretically or experimentally.

Fig. 4a and b show the iR-corrected linear sweep voltammetry (LSV) curves in 0.5 M $H_2SO_4$ and 1 M KOH, respectively, for the as-produced electrodes before and after Li-TFSI treatment (samples hereafter named $NbS_2$ and Li-TFSI-treated $NbS_2$, respectively). In addition, the LSV curves obtained for the $NbS_2$ electrode after 1000 CV cycles (samples named $NbS_2$-CV @ 1000 cycles), the commercial Pt/C benchmark and the SWCNTs (catalyst support) are also plotted. In 0.5 M $H_2SO_4$, Li-TFSI-treated $NbS_2$ exhibits a HER activity ($\eta_{10}$ = 0.31 V) significantly higher than that of as-produced $NbS_2$ ($\eta_{10}$ = 0.42 V). The electrochemical cycling also improves the HER activity of $NbS_2$, and $NbS_2$-CV @ 1000 cycles shows an $\eta_{10}$ of 0.20 V. Similar results were also obtained in 1 M KOH, where Li-TFSI-treated $NbS_2$ shows an $\eta_{10}$ of 0.33 V, similar to that of $NbS_2$-CV @ 1000 (0.31 V), whereas the untreated as-produced $NbS_2$ displays an $\eta_{10}$ of 0.49 V. Notably, the electrochemical treatment is also effective to improve the HER activity of Li-TFSI-treated $NbS_2$ (see Fig. S6†) ($\eta_{10}$ = 0.27 V in 0.5 M $H_2SO_4$; $\eta_{10}$ = 0.28 V in 1 M KOH). As discussed in the previous section,[54,55,57] the electrochemical treatments cause a progressive self-optimizing evolution of the $NbS_2$ morphology from thick-to-thin flakes. This change of the flake morphology speeds up the electron transport towards the HER-active sites by shortening the interlayer electron-transfer pathways and facilitating the access of the aqueous protons ($H_3O^+$) to the catalytic films.[54,55,57] Moreover, in the absence of catalyst film detachment/dissolution, this morphology change progressively increases the electrochemically accessible surface area of the electrodes, as demonstrated by $C_{dl}$ measurement (Fig. S5a†). In this context, it is important to evidence that the SWCNT substrate is effective in maintaining the mechanical stability of the catalyst films during electrochemical tests.[68,119] In fact, the electrode using SWCNTs as a support did not show any macroscopic material loss, while the catalyst film deposited onto glassy carbon partially precipitated on the bottom of the cell (see Fig. S4b†).

A rigorous analysis of the HER kinetics, including the extrapolation of the Tafel slope and the exchange current ($j_0$), was not carried out in this work because ambiguous results can originate from the presence of the SWCNTs. In fact, SWCNTs have a high surface area that leads to a significant capacitive

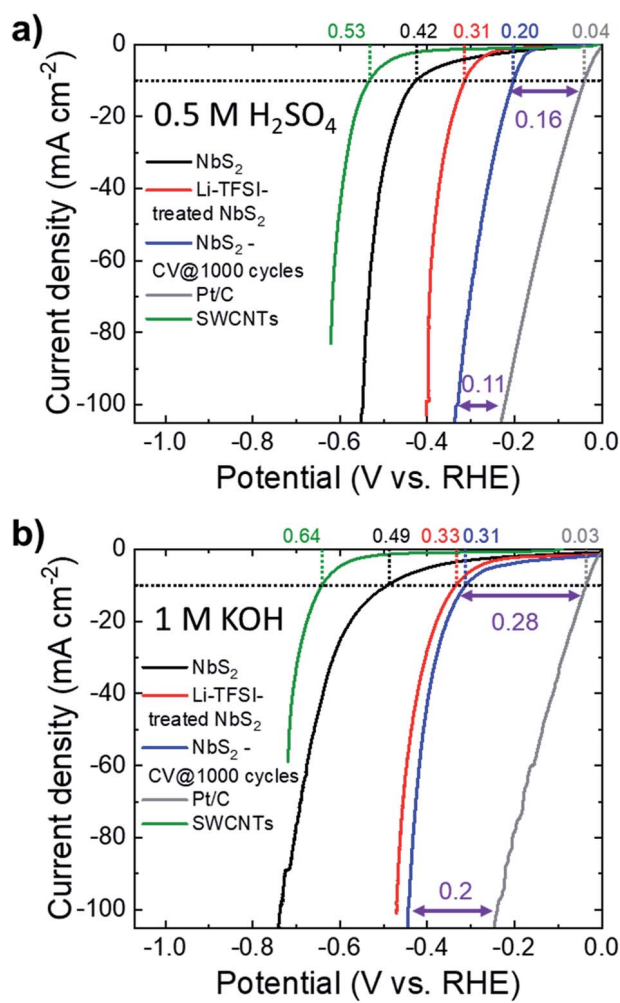

Fig. 4 (a and b) iR-corrected LSV curves for $NbS_2$, Li-TFSI-treated $NbS_2$, $NbS_2$-CV @ 1000 cycles in acidic (0.5 M $H_2SO_4$) and alkaline (1 M KOH) solutions, respectively. The LSV curves of the Pt/C benchmark and SWCNTs (catalyst support) are also shown for comparison. The $\eta_{10}$ values measured for the electrodes are also shown. For $NbS_2$-CV @ 1000 cycles, the HER overpotential vs. the Pt/C electrode at 10 mA $cm^{-2}$ and 100 mA $cm^2$ is also indicated.

current density (in the range of mA $cm^{-2}$) even at a low LSV sweep voltage rate (i.e., ≤5 mV $s^{-1}$).[51] This can be misleading during the interpretation of HER kinetic parameters.[136,137] Interestingly, the overpotentials vs. Pt/C of our electrodes at 100 mA $cm^{-2}$ are significantly lower (e.g., by ~31% and 29% for $NbS_2$-CV @ 1000 in 0.5 M $H_2SO_4$ and 1 M KOH, respectively) than those at a current density of 10 mA $cm^{-2}$. This agrees with a HER activity also relying on highly abundant basal planes[52–57] beyond that of the chalcogen and metal edges.[55,56] Noteworthily, this is substantially different from group-6 $MX_2$, in which the HER activity is restricted to only edge/defective sites.[39–43]

### Heterogeneous $NbS_2$:$MoSe_2$ catalysts

To further exploit the potential of $NbS_2$ flakes for the HER by taking advantage of the theoretically calculated database for the catalytic properties of 2D $MX_2$ (ref. 39 and 56) (and combination thereof),[56] heterogeneous electrocatalysts were produced by simply mixing LPE-produced $NbS_2$ and $MoSe_2$ flakes (electrocatalyst hereinafter referred to as $NbS_2$:$MoSe_2$). In fact, recent DFT calculations and AIMD simulations[56] have shown that the heterogeneous stacking of $NbS_2$ and (2H) $MoSe_2$ flakes tunes the $\Delta G_H^0$ of the resulting heterogeneous configurations in acidic media to optimal thermo-neutral values (*i.e.*, 0 eV) for both edge and basal sites. In more detail, the stacking process initiates an electron transfer from the $MoSe_2$ to $NbS_2$ flakes (Fig. 5a), decreasing the positive $\Delta G_H^0$ of the $NbS_2$ basal plane (theoretical values: 0.061 eV for H-coverage = 8.3% and 0.11 for H-coverage = 16.6% from ref. 138; 0.31 eV from ref. 56; 0.12 eV from ref. 39; 0.15 eV from ref. 139) to nearly 0 eV (Fig. 5b).[56]

In addition, we recently demonstrated that $MoSe_2$ flakes are also efficient HER catalysts,[28,50,51] showing (after appropriate physical or chemical treatments) an $\eta_{10}$ < 0.10 V in both acidic and alkaline solutions.[28,50,51] Theoretical calculations have proposed that the HER activity of $MoSe_2$ flakes relies on edge sites (although it is still up for debate whether they are Mo or Se-edge sites, or both),[39,56] whereas basal planes are inert ($\Delta G_H^0$ > 2 eV),[39,56] also when stacked onto $NbS_2$ flakes ($\Delta G_H^0 \sim 0.98$ eV). As shown in Fig. 5c, the alleged HER activity of the Se-edge sites in $MoSe_2$ flakes makes the latter advantageous over $MoS_2$ flakes, whose S-edges strongly bind the H, resulting in negative $\Delta G_H^0$ ($\sim$−0.45 eV).[39,56] The $\Delta G_H^0$ of the chalcogen edges of $MoSe_2$ flakes ($\sim$−0.05 eV from ref. 39 and 56) is slightly varied around the thermoneutral value when $MoSe_2$ flakes are stacked onto the $NbS_2$ flakes ($\Delta G_H^0 \sim 0.06$ eV). In addition, by stacking $MoSe_2$ flakes onto the $NbS_2$ flakes, the positive $\Delta G_H^0$ of metallic edges of $MoSe_2$ (*e.g.*, 0.31 eV from ref. 56) are low-shifted toward slightly negative near zero values (−0.01 eV from ref. 56). Similar effects have been also theoretically predicted and experimentally confirmed for $MoS_2$-coated $NbS_2$ catalysts.[109] Noteworthily, the preparation of heterogeneous catalysts from LPE-produced catalysts is a scalable approach, which does not resort to any morphological/structural and/or chemical modifications of the

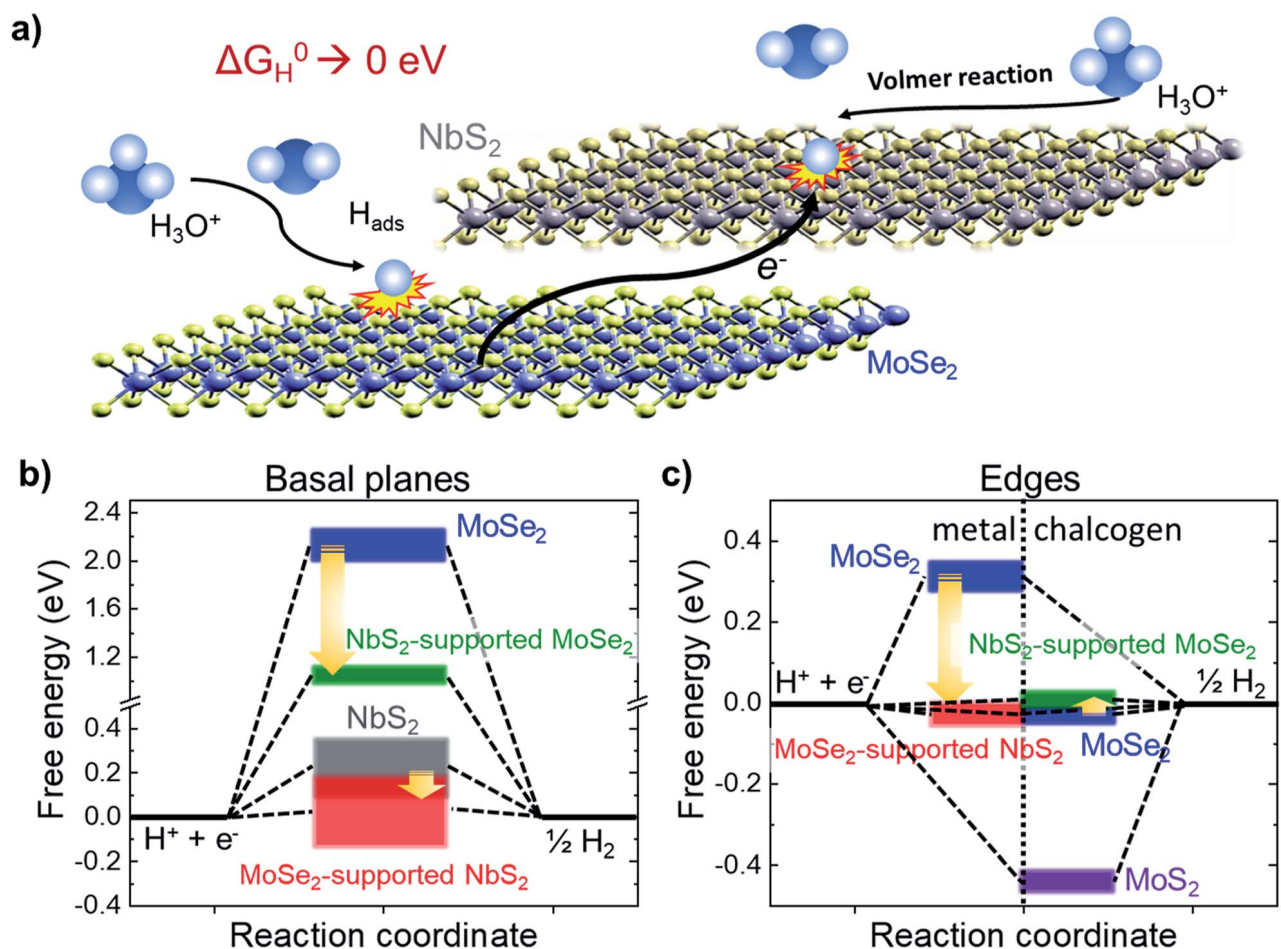

Fig. 5 (a) Schematic illustration of the HER activation of the basal plane of the $NbS_2$ flakes and metallic edges of $MoSe_2$ through stacking the $NbS_2$ and $MoSe_2$ flakes. The interaction between the flakes promotes an electron transfer from the $MoSe_2$ flakes towards the $NbS_2$ flakes, decreasing the positive $\Delta G_H^0$ values of the basal plane of the $NbS_2$ flakes and the metallic edges of $MoSe_2$ toward 0 eV, promoting the Volmer reaction ($H_3O^+$ + $e^-$ ⇌ $H_{ads}$ + $H_2O$ in acidic condition), *i.e.*, the first step of the HER. (b) Evolution of $\Delta G_H^0$ for the basal planes of the $NbS_2$ and $MoSe_2$ flakes, and the $NbS_2$:$MoSe_2$ hybrid catalysts. The data ranges have been extrapolated from literature databases (ref. 21, 39, 103 and 104). (c) Evolution of $\Delta G_H^0$ for the metallic (left side) and chalcogen (right side) edges of the $MoSe_2$ flakes, and the $NbS_2$:$MoSe_2$ hybrid catalysts. The data ranges have been extrapolated from literature databases (ref. 21 and 39).

LPE-produced flakes, such as the introduction of artificial defects, chemical doping of heteroatoms or strain constraints.[56]

The details regarding the production of $MoSe_2$ flakes are reported in the Experimental section. The morphological, structural, and chemical characterization of the LPE-produced $MoSe_2$ flakes has been recently reported by our studies.[28,50,51] In particular, the $MoSe_2$ flakes have a lateral size (measured by TEM) and thickness (measured by AFM) following lognormal distributions peaked at ~30.0 nm and ~2.9 nm, respectively.[28,50,51] Optical absorbance, Raman spectroscopy and XRD measurements have shown that $MoSe_2$ flakes optimally preserve the $MoSe_2$ bulk crystallinity.[28,50,51] In addition, the analysis of HRTEM images of the $MoSe_2$ flakes and the corresponding FFTs confirmed that the flakes display a hexagonal phase,[28] in agreement with XRD and Raman analyses.[28,50,51]

Following the protocols previously adopted for $NbS_2$ electrodes, $MoSe_2$ and $NbS_2:MoSe_2$ (material mass ratio of 1 : 1) electrodes were produced by vacuum filtration deposition of their dispersions on top of the SWNTs. Fig. S7† shows the top-view SEM images of the $MoSe_2$ and $NbS_2:MoSe_2$ electrodes, respectively. The as-produced electrodes display a surface uniformly covered by the flakes. In particular, the heterogeneous electrode (i.e., $NbS_2:MoSe_2$) shows a morphology resembling those of the $MoSe_2$ electrode, whose flakes are smaller than $NbS_2$ flakes, in agreement with the characterization of the materials (see Fig. 1g for $NbS_2$ flakes and ref. 28 and 51 for $MoSe_2$ flakes). Although it is reasonable to assume that the vacuum filtration of a mixture of $NbS_2$ and $MoSe_2$ flakes in dispersion intrinsically promotes the stacking between flakes of different materials, SEM-EDS measurements were carried out to prove at least the absence of single material domains. The cross-sectional SEM-EDS analysis of the heterogeneous films composed of $NbS_2$ and $MoSe_2$ flakes (Fig. S8a†) shows homogeneous distributions for both Nb and Mo elements, which are compatible with an optimal mixing in the dispersion used for the preparation of the electrodes. The Li-TFSI treatment of the $MoSe_2$ and $NbS_2:MoSe_2$ electrodes is expected to cause the same effects observed for the $NbS_2$ electrodes (i.e., hygroscopicity and nanostructuring of the electrodes). The SEM-EDS analysis of the Li-TFSI-treated heterogeneous films (Fig. S8b†) leads to similar results to those of the untreated film (i.e., a homogeneous distribution of both Nb and Mo elements), suggesting a similar packing between $NbS_2$ and $MoSe_2$ flakes. The top-view SEM-EDS analyses of the untreated and the Li-TFSI heterogeneous films (Fig. S8c and d,† respectively) also show homogeneous distributions of the composing elements, further confirming the absence of single material domains. Fig. 6a and b show the LSV curves in 0.5 M $H_2SO_4$ and 1 M KOH, respectively, for the heterogeneous electrodes before and after the Li-TFSI treatment (electrodes named $NbS_2:MoSe_2$ and Li-TFSI-treated $NbS_2$:$MoSe_2$). In addition, we also plotted the LSV curves obtained for the untreated heterogeneous electrode after 1000 CV cycles ($NbS_2:MoSe_2$-CV @ 1000 cycles), the commercial Pt/C benchmark and the references $NbS_2$ and $MoSe_2$. The electrochemical characterization of the electrode based on solely $MoSe_2$ before and after Li-TFSI treatment, as well as after 1000 CV cycles, is reported in Fig. S9,† showing an improvement of the initial HER

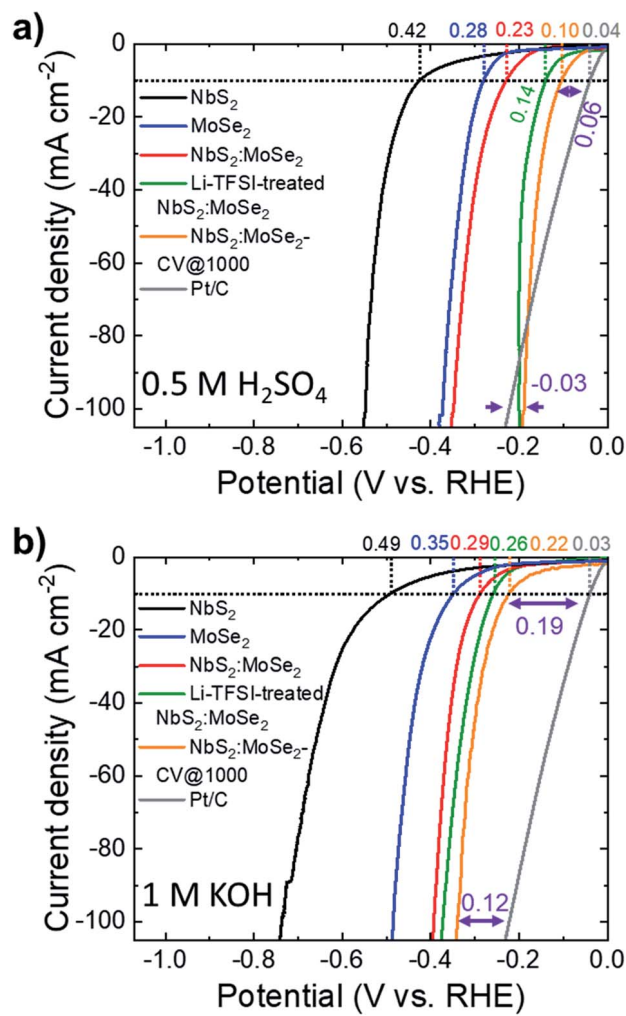

Fig. 6 (a and b) iR-corrected LSV curves for $NbS_2:MoSe_2$, Li-TFSI-treated $NbS_2:MoSe_2$ and $NbS_2$-CV @ 1000 cycles in acidic (0.5 M $H_2SO_4$) and alkaline (1 M KOH) solutions, respectively. The LSV curves of the Pt/C benchmark and $NbS_2$ and $MoSe_2$ references are also shown for comparison. The $\eta_{10}$ values measured for the electrodes are also shown. For $NbS_2:MoSe_2$-CV @ 1000 cycles, the HER overpotential vs. the Pt/C electrode at 10 mA $cm^{-2}$ and 100 mA $cm^2$ is indicated.

activity ($\eta_{10}$ = 0.28 and 0.35 V in 0.5 M $H_2SO_4$ and 1 M KOH, respectively) after both Li-TFSI treatment ($\eta_{10}$ = 0.27 and 0.30 V in 0.5 M $H_2SO_4$ and 1 M KOH, respectively) and electrochemical cycling ($\eta_{10}$ equal to 0.18 and 0.3 V in 0.5 M $H_2SO_4$ and 1 M KOH, respectively). However, the beneficial effects of the electrode treatments are less pronounced compared to the case of $NbS_2$ electrodes. The heterogeneous $NbS_2:MoSe_2$ electrodes show a HER activity ($\eta_{10}$ = 0.23 and 0.29 V in 0.5 M $H_2SO_4$ and 1 M KOH, respectively) which is higher than that of both $NbS_2$ and $MoSe_2$ electrodes. As reported for the individual counterpart-based electrodes, both the Li-TFSI treatment and the electrochemical cycling increase the HER activity of the heterogeneous electrodes. In 0.5 M $H_2SO_4$, the Li-TFSI-treated $NbS_2:MoSe_2$ electrode reaches an $\eta_{10}$ of 0.14 V, approaching that of $NbS_2:MoSe_2$-CV @ 1000 cycles ($\eta_{10}$ = 0.10 V). A similar trend was observed in 1 M KOH, in which the Li-TFSI-treated $NbS_2:MoSe_2$ electrode shows an $\eta_{10}$ of 0.26 V, whereas NbS$_2$:MoSe$_2$-CV @ 1000 cycles achieves an $\eta_{10}$ of 0.22 V. At a current density of 100 mA cm$^{-2}$, NbS$_2$:MoSe$_2$-CV @ 1000 cycles displays a HER activity higher than that of the Pt/C benchmark in an acidic environment. Under alkaline conditions, the HER overpotential vs. the Pt/C benchmark (0.12 V) at 100 mA cm$^{-2}$ is reduced by 0.03 V compared to that measured at 10 mA cm$^{-2}$ (0.19 V).

Beyond the electrocatalytic activity, the long-term steady-state durability of the catalysts is another important criterion for their practical utilization. Fig. 7a shows the chronoamperometric measurements for the Li-TFSI-treated NbS$_2$:MoSe$_2$ electrodes at a fixed potential corresponding to an initial cathodic current density of 80 mA cm$^{-2}$. In 0.5 M H$_2$SO$_4$, the electrodes retained 97.3% of the initial current densities after 12 h. In 1 M KOH, the electrode improves the HER activity over time, increasing its initial current density by +22.2%. Fig. 7b shows the iR-corrected LSV curves of the electrodes measured after the stability test, showing an $\eta_{10}$ of 0.11 and 0.12 V in 0.5 M H$_2$SO$_4$ and 1 M KOH, respectively.

Under acidic conditions, the stability of the electrode is further verified by XPS analysis (Fig. S10a†), which does not show relevant chemical alterations of the NbS$_2$ and MoSe$_2$ flakes after the long-term test. Noteworthily, these results have been achieved without using any binder, such as Nafion, which could prospectively improve further the mechanical stability of the electrodes during H$_2$ evolution.[140,141] In fact, although the gas evolution, causing the nanostructuring of the electrodes, has been correlated with the self-optimization behaviour of catalytic group-5 MX$_2$ in acidic media,[54,55,57,77] it may imply a loss of catalytic materials, which should be controlled for practical purposes. In this context, the Li-TFSI-treated electrodes perform similarly to the electrochemically cycled electrodes, but do not require in operando pre-conditioning of the electrodes. Therefore, our approach to chemically treat the NbS$_2$-based electrodes could be beneficial for controlling the design of the group-5 MX$_2$ catalyst, as well as for their long-term durability. Under alkaline conditions, the remarkable decrease of $\eta_{10}$ (from 0.26 to 0.12 V) is still under investigation. In fact, the XPS (Fig. S10†) data reveal a significant decrease of the NbS$_2$-related components (i.e., those attributed to Nb(4+) and S 2p). These observations indicate a progressive oxidation and a possible dissolution of the NbS$_2$ flakes. The decrease of the Nb atomic content relative to that of Mo is further verified by SEM-EDS analysis with a clear intensity decrease of the Nb (K$\alpha$) peak compared to the Mo (K$\alpha$) peak (Fig. S11†). Moreover, the peaks of the doublet assigned to such Nb(5+) ($\sim$207.4 and $\sim$210.1 eV) are located at binding energy slightly lower than those of the doublet assigned to Nb$_2$O$_5$ in the as-produced NbS$_2$ flakes (typically at $\sim$207.7 eV and $\sim$210.4 eV,[112–116] see Fig. S1†). Therefore, such XPS peaks could be ascribed to either sub-stoichiometric Nb$_2$O$_{5-x}$ or hydr(oxy)oxide species. Noteworthily, the latter can synergistically interact with MoSe$_2$ flakes (which are chemically stable, see Fig. S10† and ref. 28 and 51), increasing the HER activity compared to that of the initial electrode (see Fig. 7a). In fact, it has been recently demonstrated that the presence of transition metal oxides (or hydroxides) on an MX$_2$ surface can also increase the HER activity of the pristine MX$_2$ under alkaline conditions,[28,51,142] similarly to what was observed in noble metal-based electrocatalysts.[143–145]

## Conclusion

In summary, we have produced phase-mixed NbS$_2$ flakes (i.e., 1H-, 2H- and 3R-NbS$_2$ flakes) by an environmentally friendly liquid phase exfoliation (LPE) of synthetized NbS$_2$ crystals in 2-propanol (IPA). On the basis of a literature database for theoretically calculated Gibbs free energy of adsorbed atomic H ($\Delta G^0_H$), NbS$_2$ flakes have been investigated as efficient HER catalyst candidates. Vacuum filtration has been used as a scalable approach to manufacture electrodes based on NbS$_2$ flakes. The as-produced electrodes were also chemically treated with a Li-TFSI bath in order to vertically orient the NbS$_2$ flakes and to induce the nanostructuring of their morphology, increasing the water accessibility to their surface. The full potential of NbS$_2$

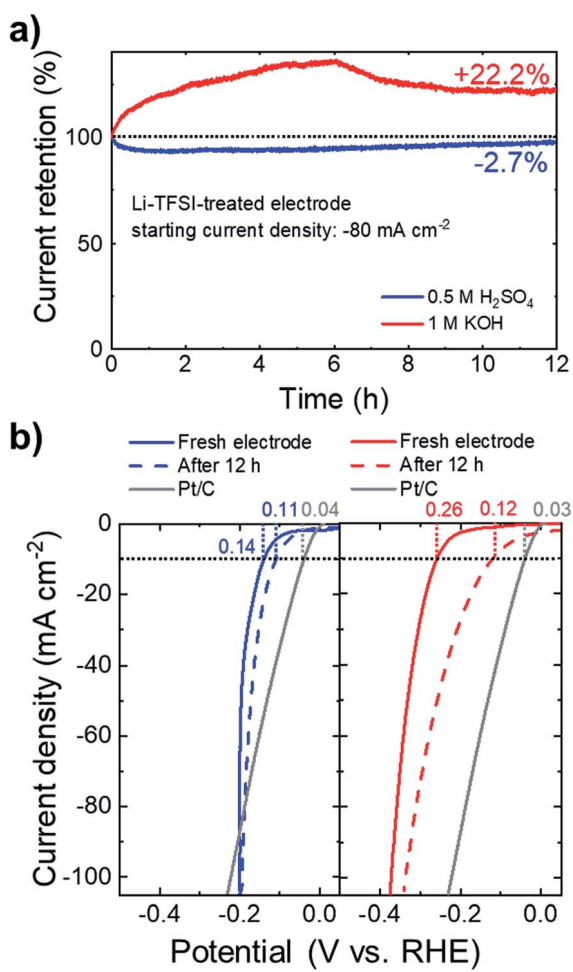

Fig. 7 (a) Chronoamperometry measurement of Li-TFSI-treated NbS$_2$:MoSe$_2$ in 0.5 M H$_2$SO$_4$ (blue line) and 1 M KOH (red line). The percentage increase of the starting cathodic current density (80 mA cm$^{-2}$) after 12 h is also indicated. (b) The LSV curves of Li-TFSI-treated NbS$_2$:MoSe$_2$ in 0.5 M H$_2$SO$_4$ (left panel) and 1 M KOH (red panel). The LSV curves of the PT/C benchmark are shown for comparison. The $\eta_{10}$ values measured for the electrodes are also indicated.

flakes was exploited in a heterogeneous electrode obtained by mixing NbS$_2$ and MoSe$_2$ flakes. In fact, the flake stacking increases the HER activity of both the basal plane of the NbS$_2$ flakes and the metallic edge of MoSe$_2$, without affecting that of the catalytic sites of the single counterparts. In acidic solution (0.5 M H$_2$SO$_4$), the designed NbS$_2$-based electrocatalysts reach an overpotential at a cathodic current density of 10 mA cm$^{-2}$ ($\eta_{10}$) as low as 0.14 and 0.10 V after chemical Li-TFSI treatment and electrochemical cycling, respectively. At current density $\geq$100 mA cm$^{-2}$ the HER activity of the electrodes is higher than that of the Pt/C benchmark (overpotential reduced by 0.03 V). For the first time, the HER activity of NbS$_2$-based electrocatalysts is investigated in alkaline solutions (1 M KOH), showing a promising $\eta_{10}$ of 0.26 and 0.22 V after chemical Li-TFSI treatment and electrochemical cycling, respectively. The durability of the electrodes is validated over 12 h of continuous operation at a fixed potential corresponding to an initial cathodic current density of 80 mA cm$^{-2}$. The results show an optimal performance retention of the chemically treated electrodes in 0.5 M H$_2$SO$_4$ (97.3%). In 1 M KOH, the electrodes significantly improve their initial cathodic current density (+22.2%), displaying an $\eta_{10}$ of only 0.12 V after 12 h. However, chemical changes affect the NbS$_2$ flakes during HER operation, and further studies are needed to validate practical applications. Our results provide new insight into the exploitation of metallic group-5 TMDs for the HER under pH-universal conditions via scalable material preparation and electrode processing. The chemical treatment of the electrodes may be advantageous to their in operando electrochemical cycling, because it can be prospectively coupled with the use of proton conducting binders to avoid catalyst dissolution/loss.

## Experimental

### Materials

Nb (99.9%, <100 μm) and S (99.999%, <6 mm) powders were purchased from Strem, USA. H$_2$SO$_4$ (99.999%), KOH ($\geq$85% purity, ACS reagent, pellets), Pt/C (10 wt% loading) and Nafion solution (5 wt%) were purchased from Sigma Aldrich. The SWCNTs (>90% purity) were purchased from Cheap Tubes. The MoSe$_2$ crystal was purchased from HQ graphene.

### Crystals' synthesis and exfoliation

NbS$_2$ crystals were made by direct synthesis from elements. An amount of Nb and S powders with a Nb : S stoichiometry of ca. 1 : 2 (1 wt% excess of S) and corresponding to a total mass of 10 g of NbS$_2$ was placed in a quartz glass ampoule (20 mm × 120 mm) and sealed under high vacuum (1 × 10$^{-3}$ Pa). The ampoule was heated at 450 °C for 12 h and subsequently at 600 °C for 48 h. Finally the derived product was treated at 900 °C for 48 h and cooled down to room temperature over a period of 24 h, obtaining NbS$_2$ crystal powder. The NbS$_2$ and MoSe$_2$ flakes were produced by LPE in IPA from the as-synthesized NbS$_2$ crystal powder and purchased MoSe$_2$ crystal,[63,80] followed by SBS to remove the unexfoliated material by ultracentrifugation.[81,82] Experimentally, 50 mg of bulk crystals were added to 50 mL of anhydrous IPA and ultrasonicated in a bath sonicator (Branson® 5800 cleaner, Branson Ultrasonics) for 6 h. The resulting dispersions were ultracentrifuged at 2700 g (Optima™ XE-90 with a SW32Ti rotor, Beckman Coulter) for 20 min at 15 °C in order to separate un-exfoliated bulk crystals (collected as sediment) from the exfoliated materials that remained in the supernatant. Then, 80% of the supernatant was collected by pipetting, obtaining an exfoliated material dispersion. The concentration of the NbS$_2$ and MoSe$_2$ flake dispersions was 0.30 g L$^{-1}$ and 0.25 g L$^{-1}$, respectively.

### Material dispersions' preparation

The dispersions of NbS$_2$ and MoSe$_2$ flakes were used as-produced. The heterogeneous dispersions of the NbS$_2$ and MoSe$_2$ flakes were obtained by mixing the single material component dispersions with a material weight ratio (w/w) of 1 : 1. The SWCNT dispersions were produced by dispersing SWCNTs in N-methyl-2-pyrrolidone (NMP) with a concentration of 0.2 g L$^{-1}$ by means of ultrasonication-based de-bundling.[146–148] In particular, 10 mg of SWCNT powder was added to 50 mL of NMP. The dispersion was then ultrasonicated for 30 min by using a sonic tip (Vibra-cell 75185, Sonics) with the vibration amplitude set to 45%. The sonic tip was pulsed for 5 s on and 2 s off to reduce the solvent heating, which was also mitigated by using an ice bath around the vessel. The dispersion of Pt/C was produced by dissolving 4 mg of Pt/C and 80 μL of Nafion solution in 1 mL of 1 : 4 v/v ethanol/water.

### Materials' characterization

The SEM analysis of the as-synthesized crystals was carried out using a JEOL JSM-7500FA equipped with a cold FEG, operated at 10 kV accelerating voltage. The EDS analyses were performed using an Oxford X-Max 80 system equipped with an 80 mm$^2$ silicon drift detector (SDD) and using Oxford's AZtec TruMap software. The displayed SEM-EDS maps were obtained by background subtraction and peak deconvolution in order to correct for the overlapping S (Kα peak) and Nb (Lβ peak) features in the EDS spectrum. The surface of the samples was cleaned by mechanical exfoliation and imaged without any conductive coating. The BF-TEM images of the NbS$_2$ flakes were acquired with a JEM 1011 (JEOL) TEM (thermionic W filament), operated at 100 kV. The morphological and statistical analysis was performed by using ImageJ software (NIH) and OriginPro 9.1 software (OriginLab), respectively. The samples for the TEM measurements were prepared by drop casting the as-prepared exfoliated material dispersions onto ultrathin C-on-holey C-coated Cu grids and rinsed with deionized water and subsequently dried overnight under vacuum. The HRTEM, HAADF-STEM imaging and STEM-EDS analyses were carried out on an image-Cs-corrected JEOL JEM-2200FS TEM (Schottky emitter), operated at 200 kV, equipped with an in-column image filter (Ω-type) and a Bruker XFlash 5060 EDS detector. For these analyses, the NbS$_2$ flake dispersion was drop cast onto a holey-carbon-coated Cu grid. The displayed STEM-EDS maps are "quantitative", in analogy to what was done for SEM-EDS maps of the crystals.

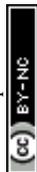

The AFM images were acquired with a Nanowizard III (JPK Instruments, Germany), mounted on an Axio Observer D1 (Carl Zeiss, Germany) inverted optical microscope. The AFM measurements were carried out by means of PPP-NCHR cantilevers (Nanosensors, USA) having a nominal tip diameter of 10 nm. We used a drive frequency of ∼295 kHz. We collected intermittent contact mode AFM images (512 × 512 data points) of 2.5 × 2.5 μm$^2$ by keeping the working set point above 70% of the free oscillation amplitude. We used a scan rate of 0.7 Hz for the acquisition of the images. JPK Data Processing software (JPK Instruments, Germany) was exploited to process the height profiles, while the data were analysed by using OriginPro 9.1 software. The latter was also used to carry out the statistical analysis on multiple AFM images for all the tested samples. The samples were prepared by drop-casting the as-prepared exfoliated material dispersions onto mica sheets (G250-1, Agar Scientific Ltd.) and dried overnight under vacuum. The XRD measurements were acquired with a PANalytical Empyrean using Cu K$_\alpha$ radiation. The samples for XRD were prepared by drop-casting the as-prepared exfoliated material dispersions onto Si/SiO$_2$ substrates and dried overnight under vacuum. Raman spectroscopy measurements were performed by using a Renishaw microRaman inVia 1000 mounted with a 50× objective, with an excitation wavelength of 532 nm and an incident power of 1 mW on the samples. For each sample, we collected 50 spectra. The samples were prepared by drop casting the as-prepared exfoliated material dispersions onto Si/SiO$_2$ substrates and subsequently dried under vacuum. The XPS characterization was performed with a Kratos Axis UltraDLD spectrometer, having a monochromatic Al Kα source (15 kV, 20 mA). The spectra were acquired on a 300 × 700 μm$^2$ area. A constant pass energy of 160 eV and an energy step of 1 eV were used to collect wide scans. High-resolution spectra were acquired at a constant pass energy of 10 eV with an energy step of 0.1 eV. The binding energy scale was referenced to the C 1s peak at 284.8 eV. The spectra were analysed using CasaXPS software (version 2.3.17). The samples were prepared by drop casting the NbS$_2$ flake dispersions onto a Si/SiO$_2$ substrate (LDB Technologies Ltd), followed by a drying process under vacuum.

### Electrodes' fabrication

The electrodes were produced by sequentially depositing the SWCNT and exfoliated catalytic material (NbS$_2$, MoSe$_2$ and heterogeneous NbS$_2$:MoSe$_2$) dispersions onto nylon membranes (Whatman® nylon membrane filters, 0.2 μm pore size) through a vacuum filtration process.[48] The material mass loading was 1.31 mg cm$^{-2}$ for SWCNTs and 0.2 mg cm$^{-2}$ for the exfoliated catalytic materials. The electrode area was 3.8 cm$^2$. Additional electrodes were produced by depositing the as-produced NbS$_2$ flake dispersion onto glassy carbon by a drop casting method (NbS$_2$ flake mass loading = 0.5 mg cm$^{-2}$). Before the electrochemical characterization, the as-prepared electrodes were dried overnight at room temperature. The electrodes were chemically treated with Li-TFSI inside a glove box. Experimentally, 2 mg of Li-TFSI (Sigma-Aldrich) were dissolved in 10 mL of anhydrous acetonitrile (ACN) (Sigma-Aldrich) and stirred for 15 min at room temperature to make a 0.2 g L$^{-1}$ solution. Subsequently, the electrodes were immersed in the Li-TFSI solution in a closed vial for 20 min at a temperature of 70 °C. The electrodes were removed (without rinsing) and subsequently annealed at 100 °C for 5 min in air in order to remove the residuals of ACN. Electrodes made entirely of SWCNTs were also produced as a reference with a mass loading of 1.31 mg cm$^{-2}$. A Pt/C electrode was produced as a benchmark for the HER by depositing the corresponding dispersion onto glassy carbon substrates (Sigma Aldrich). The active material mass loading of the electrodes was 0.262 mg cm$^{-2}$ for Pt/C, in agreement with previously reported protocols.[28,51]

### Electrode characterization

The XPS analyses of the as-produced electrodes and the electrodes after chemical and electrochemical treatments were performed with the same set-up and parameters used for the characterization of the materials. The detailed SEM images of the electrodes were collected on a Helios Nanolab® 600 Dual-Beam microscope (FEI Company) and 5 kV and 0.2 nA as measurement conditions. The SEM-EDS measurements were performed on a JEOL JSM-6490LA SEM at 30 kV. The electrochemical measurements were performed at room temperature in a flat-bottom fused silica cell using the three-electrode configuration of a potentiostat/galvanostat station (VMP3, Biologic), controlled via own software. A glassy carbon rod and saturated KCl Ag/AgCl were employed as the counter electrode and reference electrode, respectively. The measurements were carried out in 200 mL of 0.5 M H$_2$SO$_4$ or 1 M KOH. Before starting the measurements, the oxygen was purged from the electrolyte by flowing N$_2$ gas throughout the liquid volume using a porous frit. The Nernst equation: $E_{RHE} = E_{Ag/AgCl} + 0.059 \times pH + E^0_{Ag/AgCl}$, where $E_{RHE}$ is the converted potential vs. RHE, $E_{Ag/AgCl}$ is the experimental potential measured against the Ag/AgCl reference electrode, and $E^0_{Ag/AgCl}$ is the standard potential of Ag/AgCl at 25 °C (0.1976 V vs. RHE), was used to convert the potential difference between the working electrode and the Ag/AgCl reference electrode to the reversible hydrogen electrode (RHE) scale. The double-layer capacitances ($C_{dl}$) of the untreated and treated NbS$_2$ films deposited onto glassy carbon were estimated by cyclic voltammetry (CV) measurements in a non-faradaic region of potential (between 0.2 and 0.4 V vs. RHE) at various potential scan rates (ranging from 20 to 600 mV s$^{-1}$). Cyclic voltammetry was also exploited for the electrochemical treatment of the electrodes. Experimentally, 1000 CV scans were carried out between 0.25 and 0.75 V vs. RHE at a potential scan rate of 100 mV s$^{-1}$. The LSV curves of the electrodes using a SWCNT substrate were acquired at a 5 mV s$^{-1}$ scan rate and were iR-corrected by considering i as the measured working electrode current and R as the series resistance arising from the working electrode substrate and electrolyte resistances. Electrochemical impedance spectroscopy (EIS) measurements were performed on the electrodes at open circuit potential and at a frequency of 10 kHz to measure R (in agreement with methods reported in previous studies).[48,50,51] The stability tests were performed by chronoamperometry measurements, i.e., by measuring the current in potentiostatic mode over time, at a fixed overpotential corresponding to a cathodic current density of 80 mA cm$^{-2}$.

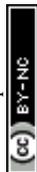

## Conflicts of interest

There are no conflicts to declare.

## Acknowledgements

This project has received funding from the European Union's Horizon 2020 research and innovation program under grant agreement No. 785219—GrapheneCore2. We thank Lea Pasquale and Marco Salerno (Materials Characterization Facility – Istituto Italiano di Tecnologia) for support in XRD and AFM data acquisition; the Electron Microscopy Facility – Istituto Italiano di Tecnologia for support in TEM and SEM/EDS data acquisition; and the Clean Room facility – Istituto Italiano di Tecnologia for the access to carry out part of the SEM characterization. This work was supported by the project Advanced Functional Nanorobots (reg. No. CZ.02.1.01/0.0/0.0/15_003/0000444 financed by the EFRR). Z. S and V. M. were supported by the Czech Science Foundation (GACR No. 17-11456S).

## Notes and references

1 B. Johnston, M. C. Mayo and A. Khare, *Technovation*, 2005, **25**, 569–585.
2 M. Momirlan and T. N. Veziroglu, *Int. J. Hydrogen Energy*, 2005, **30**, 795–802.
3 M. S. Dresselhaus and I. L. Thomas, *Nature*, 2001, **414**, 332.
4 N. S. Lewis and D. G. Nocera, *Proc. Natl. Acad. Sci. U. S. A.*, 2006, **103**, 15729–15735.
5 J. A. Turner, *Science*, 2004, **305**, 972–974.
6 W. E. Winsche, K. C. Hoffman and F. J. Salzano, *Science*, 1973, **180**, 1325–1332.
7 M. K. Debe, *Nature*, 2012, **486**, 43.
8 K. E. Ayers, J. N. Renner, N. Danilovic, J. X. Wang, Y. Zhang, R. Maric and H. Yu, *Catal. Today*, 2016, **262**, 121–132.
9 J. Zheng, W. Sheng, Z. Zhuang, B. Xu and Y. Yan, *Sci. Adv.*, 2016, **2**, e1501602.
10 S. Enthaler, K. Junge and M. Beller, *Angew. Chem., Int. Ed.*, 2008, **47**, 3317–3321.
11 D. A. Singer, *Econ. Geol.*, 1995, **90**, 88–104.
12 C. Sealy, *Mater. Today*, 2008, **12**, 65–68.
13 M. Ball and M. Weeda, *Int. J. Hydrogen Energy*, 1995, **90**, 88–104.
14 P. Wang, K. Jiang, G. Wang, J. Yao and X. Huang, *Angew. Chem., Int. Ed.*, 2016, **55**, 12859–12863.
15 Z. Cao, Q. Chen, J. Zhang, H. Li, Y. Jiang, S. Shen, G. Fu, B. Lu, Z. Xie and L. Zheng, *Nat. Commun.*, 2017, **8**, 15131.
16 W. F. Chen, K. Sasaki, C. Ma, A. I. Frenkel, N. Marinkovic, J. T. Muckerman, Y. Zhu and R. R. Adzic, *Angew. Chem., Int. Ed.*, 2012, **51**, 6131–6135.
17 L. Yu, S. Song, B. McElhenny, F. Ding, D. Luo, Y. Yu, S. Chen and Z. Ren, *J. Mater. Chem. A*, 2019, **7**, 19728–19732.
18 J. X. Feng, J. Q. Wu, Y. X. Tong and G. R. Li, *J. Am. Chem. Soc.*, 2018, **140**, 610–617.
19 Y. Hou, M. R. Lohe, J. Zhang, S. Liu, X. Zhuang and X. Feng, *Energy Environ. Sci.*, 2016, **9**, 478–483.
20 Y. Hou, M. Qiu, T. Zhang, X. Zhuang, C.-S. Kim, C. Yuan and X. Feng, *Adv. Mater.*, 2017, **29**, 1701589.
21 E. Petroni, E. Lago, S. Bellani, D. W. Boukhvalov, A. Politano, B. Gürbulak, S. Duman, M. Prato, S. Gentiluomo, R. Oropesa-Nuñez, J.-K. Panda, P. S. Toth, A. E. Del Rio Castillo, V. Pellegrini and F. Bonaccorso, *Small*, 2018, **14**, 1800749.
22 Y. Hou, M. Qiu, G. Nam, M. G. Kim, T. Zhang, K. Liu, X. Zhuang, J. Cho, C. Yuan and X. Feng, *Nano Lett.*, 2017, **17**, 4202–4209.
23 Y. Hou, M. Qiu, M. G. Kim, P. Liu, G. Nam, T. Zhang, X. Zhuang, B. Yang, J. Cho, M. Chen, C. Yuan, L. Lei and X. Feng, *Nat. Commun.*, 2019, **10**, 1392.
24 F. H. Saadi, A. I. Carim, W. S. Drisdell, S. Gul, J. H. Baricuatro, J. Yano, M. P. Soriaga and N. S. Lewis, *J. Am. Chem. Soc.*, 2017, **139**, 12927–12930.
25 P. Xiao, M. A. Sk, L. Thia, X. Ge, R. J. Lim, J.-Y. Wang, K. H. Lim and X. Wang, *Energy Environ. Sci.*, 2014, **7**, 2624–2629.
26 C. Wan, Y. N. Regmi and B. M. Leonard, *Angew. Chem.*, 2014, **126**, 6525–6528.
27 L. Liao, S. Wang, J. Xiao, X. Bian, Y. Zhang, M. D. Scanlon, X. Hu, Y. Tang, B. Liu and H. H. Girault, *Energy Environ. Sci.*, 2014, **7**, 387–392.
28 L. Najafi, S. Bellani, R. Oropesa-Nuñez, M. Prato, B. Martín-García, R. Brescia and F. Bonaccorso, *ACS Nano*, 2019, **13**, 3162–3176.
29 J. Lai, S. Li, F. Wu, M. Saqib, R. Luque and G. Xu, *Energy Environ. Sci.*, 2016, **9**, 1210–1214.
30 L. Najafi, S. Bellani, R. Oropesa-Nuñez, B. Martín-García, M. Prato and F. Bonaccorso, *ACS Appl. Energy Mater.*, 2019, **2**, 5373–5379.
31 C. Lei, Y. Wang, Y. Hou, P. Liu, J. Yang, T. Zhang, X. Zhuang, M. Chen, B. Yang, L. Lei, C. Yuan, M. Qiu and X. Feng, *Energy Environ. Sci.*, 2019, **12**, 149–156.
32 M. Chhowalla, Z. Liu and H. Zhang, *Chem. Soc. Rev.*, 2015, **44**, 2584–2586.
33 M. Chhowalla, H. S. Shin, G. Eda, L.-J. Li, K. P. Loh and H. Zhang, *Nat. Chem.*, 2013, **5**, 263.
34 D. Kong, J. J. Cha, H. Wang, H. R. Lee and Y. Cui, *Energy Environ. Sci.*, 2013, **6**, 3553–3558.
35 J. Yang and H. S. Shin, *J. Mater. Chem. A*, 2014, **2**, 5979–5985.
36 D. Merki and X. Hu, *Energy Environ. Sci.*, 2011, **4**, 3878–3888.
37 M. Pumera, Z. Sofer and A. Ambrosi, *J. Mater. Chem. A*, 2014, **2**, 8981–8987.
38 H. Li, X. Jia, Q. Zhang and X. Wang, *Chem*, 2018, **4**, 1510–1537.
39 C. Tsai, K. Chan, J. K. Nørskov and F. Abild-Pedersen, *Surf. Sci.*, 2015, **640**, 133–140.
40 D. Er, H. Ye, N. C. Frey, H. Kumar, J. Lou and V. B. Shenoy, *Nano Lett.*, 2018, **18**, 3943–3949.

41 C. Tsai, H. Li, S. Park, J. Park, H. S. Han, J. K. Nørskov, X. Zheng and F. Abild-Pedersen, *Nat. Commun.*, 2017, **8**, 15113.
42 J. Kibsgaard, Z. Chen, B. N. Reinecke and T. F. Jaramillo, *Nat. Mater.*, 2012, **11**, 963.
43 J. Xie, H. Zhang, S. Li, R. Wang, X. Sun, M. Zhou, J. Zhou, X. W. Lou and Y. Xie, *Adv. Mater.*, 2013, **25**, 5807–5813.
44 D. McAteer, Z. Gholamvand, N. McEvoy, A. Harvey, E. O'Malley, G. S. Duesberg and J. N. Coleman, *ACS Nano*, 2016, **10**, 672–683.
45 Y. Yu, S.-Y. Huang, Y. Li, S. N. Steinmann, W. Yang and L. Cao, *Nano Lett.*, 2014, **14**, 553–558.
46 Y. Huang, Y.-E. Miao, J. Fu, S. Mo, C. Wei and T. Liu, *J. Mater. Chem. A*, 2015, **3**, 16263–16271.
47 G. Ye, Y. Gong, J. Lin, B. Li, Y. He, S. T. Pantelides, W. Zhou, R. Vajtai and P. M. Ajayan, *Nano Lett.*, 2016, **16**, 1097–1103.
48 L. Najafi, S. Bellani, B. Martín-García, R. Oropesa-Nuñez, A. E. Del Rio Castillo, M. Prato, I. Moreels and F. Bonaccorso, *Chem. Mater.*, 2017, **29**, 5782–5786.
49 T. F. Jaramillo, K. P. Jørgensen, J. Bonde, J. H. Nielsen, S. Horch and I. Chorkendorff, *Science*, 2007, **317**, 100–102.
50 L. Najafi, S. Bellani, R. Oropesa-Nuñez, A. Ansaldo, M. Prato, A. E. Del Rio Castillo and F. Bonaccorso, *Adv. Energy Mater.*, 2018, **8**, 1703212.
51 L. Najafi, S. Bellani, R. Oropesa-Nuñez, A. Ansaldo, M. Prato, A. E. Del Rio Castillo and F. Bonaccorso, *Adv. Energy Mater.*, 2018, **8**, 1801764.
52 C. Zhu, D. Gao, J. Ding, D. Chao and J. Wang, *Chem. Soc. Rev.*, 2018, **47**, 4332–4356.
53 Y. Huan, J. Shi, X. Zou, Y. Gong, Z. Zhang, M. Li, L. Zhao, R. Xu, S. Jiang, X. Zhou, M. Hong, C. Xie, H. Li, X. Lang, Q. Zhang, L. Gu, X. Yan and Y. Zhang, *Adv. Mater.*, 2018, **30**, 1705916.
54 Y. Liu, J. Wu, K. P. Hackenberg, J. Zhang, Y. M. Wang, Y. Yang, K. Keyshar, J. Gu, T. Ogitsu, R. Vajtai, J. Lou, P. M. Ajayan, B. C. Wood and B. I. Yakobson, *Nat. Energy*, 2017, **2**, 17127.
55 J. Shi, X. Wang, S. Zhang, L. Xiao, Y. Huan, Y. Gong, Z. Zhang, Y. Li, X. Zhou, M. Hong, Q. Fang, Q. Zhang, X. Liu, L. Gu, Z. Liu and Y. Zhang, *Nat. Commun.*, 2017, **8**, 958.
56 B. Han, S. H. Noh, D. Choi, M. H. Seo, J. Kang and J. Hwang, *J. Mater. Chem. A*, 2018, **6**, 20005–20014.
57 J. Zhang, J. Wu, X. Zou, K. Hackenberg, W. Zhou, W. Chen, J. Yuan, K. Keyshar, G. Gupta, A. Mohite, P. M. Ajayan and J. Lou, *Mater. Today*, 2019, **25**, 28–34.
58 C. Huang, X. Wang, D. Wang, W. Zhao, K. Bu, J. Xu, X. Huang, Q. Bi, J. Huang and F. Huang, *Chem. Mater.*, 2019, **31**, 4726–4731.
59 J. Hwang, S. H. Noh and B. Han, *Appl. Surf. Sci.*, 2019, **471**, 545–552.
60 D. N. Chirdon and Y. Wu, *Nat. Energy*, 2017, **2**, 17132.
61 Z. Cai, B. Liu, X. Zou and H. M. Cheng, *Chem. Rev.*, 2018, **118**, 6091–6133.
62 Y. Huang, E. Sutter, N. N. Shi, J. Zheng, T. Yang, D. Englund, H. J. Gao and P. Sutter, *ACS Nano*, 2015, **9**, 10612–10620.
63 V. Nicolosi, M. Chhowalla, M. G. Kanatzidis, M. S. Strano and J. N. Coleman, *Science*, 2013, **340**, 1226419.
64 A. E. Del Rio Castillo, V. Pellegrini, A. Ansaldo, F. Ricciardella, H. Sun, L. Marasco, J. Buha, Z. Dang, L. Gagliani, E. Lago, N. Curreli, S. Gentiluomo, F. Palazon, M. Prato, R. Oropesa-Nuñez, P. S. Toth, E. Mantero, M. Crugliano, A. Gamucci, A. Tomadin, M. Polini and F. Bonaccorso, *Mater. Horiz.*, 2018, **5**, 890–904.
65 F. Bonaccorso, A. Lombardo, T. Hasan, Z. Sun, L. Colombo and A. C. Ferrari, *Mater. Today*, 2012, **15**, 564–589.
66 S. Bellani, F. Wang, G. Longoni, L. Najafi, R. Oropesa-Nuñez, A. E. Del Rio Castillo, M. Prato, X. Zhuang, V. Pellegrini, X. Feng and F. Bonaccorso, *Nano Lett.*, 2018, **18**, 7155–7164.
67 S. Bellani, B. Martín-García, R. Oropesa-Nuñez, V. Romano, L. Najafi, C. Demirci, M. Prato, A. E. Del Rio Castillo, L. Marasco, E. Mantero, G. D'Angelo and F. Bonaccorso, *Nanoscale Horiz.*, 2019, **4**, 1077–1091.
68 S. Bellani, E. Petroni, A. E. Del Rio Castillo, N. Curreli, B. Martín-García, R. Oropesa-Nuñez, M. Prato and F. Bonaccorso, *Adv. Funct. Mater.*, 2019, **29**, 1807659.
69 A. Agresti, S. Pescetelli, A. L. Palma, B. Martin-Garcia, L. Najafi, S. Bellani, I. Moreels, M. Prato, F. Bonaccorso and A. Di Carlo, *ACS Energy Lett.*, 2019, **4**, 1862–1871.
70 Y. L. Zhong, Z. Tian, G. P. Simon and D. Li, *Mater. Today*, 2015, **18**, 73–78.
71 K. Parvez, S. Yang, X. Feng and K. Müllen, *Synth. Met.*, 2015, **210**, 123–132.
72 S. Bellani, L. Najafi, A. Capasso, A. E. Del Rio Castillo, M. R. Antognazza and F. Bonaccorso, *J. Mater. Chem. A*, 2017, **5**, 4384–4396.
73 S. Yang, M. R. Lohe, K. Müllen and X. Feng, *Adv. Mater.*, 2016, **28**, 6213–6221.
74 P. Yu, S. E. Lowe, G. P. Simon and Y. L. Zhong, *Curr. Opin. Colloid Interface Sci.*, 2015, **20**, 329–338.
75 R. Yan, G. Khalsa, B. T. Schaefer, A. Jarjour, S. Rouvimov, K. C. Nowack, H. G. Xing and D. Jena, *Appl. Phys. Express*, 2019, **12**, 23008.
76 S. K. Basu and M. Taniguchi, *J. Therm. Anal.*, 1987, **32**, 1105–1113.
77 K. Hackenberg, *et al.*, Self-improving electrocatalysts for gas evolution reactions, *US Pat.* 20160153098A1, 2016.
78 X. Chia, A. Ambrosi, P. Lazar, Z. Sofer and M. Pumera, *J. Mater. Chem. A*, 2016, **4**, 4241–14253.
79 M. Leroux, L. Cario, A. Bosak and P. Rodière, *Phys. Rev. B*, 2018, **97**, 195140.
80 F. Bonaccorso, A. Bartolotta, J. N. Coleman and C. Backes, *Adv. Mater.*, 2016, **28**, 6136–6166.
81 A. Capasso, A. E. Del Rio Castillo, H. Sun, A. Ansaldo, V. Pellegrini and F. Bonaccorso, *Solid State Commun.*, 2015, **224**, 53–63.
82 O. M. Maragó, F. Bonaccorso, R. Saija, G. Privitera, P. G. Gucciardi, M. A. Iatì, G. Calogero, P. H. Jones, F. Borghese, P. Denti, V. Nicolosi and A. C. Ferrari, *ACS Nano*, 2010, **4**, 7515–7523.

83 D. Zheng, X. Zhang, C. Li, M. E. McKinnon, R. G. Sadok, D. Qu, X. Yu, H. S. Lee, X. Q. Yang and D. Qu, *J. Electrochem. Soc.*, 2015, **162**, A203–A206.
84 F. Bonaccorso, A. Lombardo, T. Hasan, Z. Sun, L. Colombo and A. C. Ferrari, *Mater. Today*, 2012, **15**, 564–589.
85 S. Zhao, T. Hotta, T. Koretsune, K. Watanabe, T. Taniguchi, K. Sugawara, T. Takahashi, H. Shinohara and R. Kitaura, *2D Mater.*, 2016, **3**, 025027.
86 X. Wang, J. Lin, Y. Zhu, C. Luo, K. Suenaga, C. Cai and L. Xie, *Nanoscale*, 2017, **9**, 16607–16611.
87 Z. Li, W. Yang, Y. Losovyj, J. Chen, E. Xu, H. Liu, M. Werbianskyj, H. A. Fertig, X. Ye and S. Zhang, *Nano Res.*, 2018, **11**, 5978–5988.
88 J. A. Wilson and A. D. Yoffe, *Adv. Phys.*, 2015, **6**, 5763.
89 S. Jeong, D. Yoo, M. Ahn, P. Miro, T. Heine and J. Cheon, *Nat. Commun.*, 2015, **6**, 5763.
90 P. Rabu, A. Meerschaut, J. Rouxel and G. A. Wiegers, *J. Solid State Chem.*, 1990, **88**, 451–458.
91 W. Ge, K. Kawahara, M. Tsuji and H. Ago, *Nanoscale*, 2013, **5**, 5773–5778.
92 F. Jellinek, G. Brauers and H. Muller, *Nature*, 1960, **185**, 376–377.
93 Z.-L. Liu, L.-C. Cai and X.-L. Zhang, *J. Alloys Compd.*, 2014, **610**, 472–477.
94 C. J. Carmalt, T. D. Manning, I. P. Parkin, E. S. Peters and A. L. Hector, *J. Mater. Chem.*, 2004, **14**, 290–291.
95 C. J. Carmalt, T. D. Manning, I. P. Parkin, E. S. Peters and A. L. Hector, *Thin Solid Films*, 2004, **469–470**, 495–499.
96 Y. Ma, A. Kuc, Y. Jing, P. Philipsen and T. Heine, *Angew. Chem., Int. Ed.*, 2017, **56**, 10214–10218.
97 Y. Liao, K. S. Park, P. Xiao, G. Henkelman, W. Li and J. B. Goodenough, *Chem. Mater.*, 2013, **25**, 1699–1705.
98 Y. Liao, K. S. Park, P. Singh, W. Li and J. B. Goodenough, *J. Power Sources*, 2014, **245**, 27–32.
99 http://www.hqgraphene.com/NbS2.php, accessed on 28 August 2019.
100 http://www.hqgraphene.com/3R-NbS2.php, accessed on 28 August 2019.
101 N. Yue, J. Weicheng, W. Rongguo, D. Guomin and H. Yifan, *J. Mater. Chem. A*, 2016, **4**, 8198–8203.
102 W. G. McMullan and J. C. Irwin, *Solid State Commun.*, 1983, **45**, 557–560.
103 S. Nakashima, Y. Tokuda, A. Mitsuishi, R. Aoki and Y. Hamaue, *Solid State Commun.*, 1982, **42**, 601–604.
104 S. Onari, T. Arai, R. Aoki and S. Nakamura, *Solid State Commun.*, 1979, **31**, 577–579.
105 D. Gopalakrishnan, A. Lee, N. K. Thangavel and L. M. Reddy Arava, *Sustainable Energy Fuels*, 2017, **2**, 96–102.
106 Y. Ding, Y. Wang, J. Ni, L. Shi, S. Shi and W. Tang, *Phys. B*, 2011, **406**, 2254–2260.
107 J. K. Dash, L. Chen, P. H. Dinolfo, T. M. Lu and G. C. Wang, *J. Phys. Chem. C*, 2015, **119**, 19763–19771.
108 https://materialsproject.org/materials/mp-10033/, accessed on 28 August 2019.
109 X. Zhou, S.-H. Lin, X. Yang, H. Li, M. N. Hedhili, L.-J. Li, W. Zhang and Y. Shi, *Nanoscale*, 2018, **10**, 3444–3450.
110 H. Bark, Y. Choi, J. Jung, J. H. Kim, H. Kwon, J. Lee, Z. Lee, J. H. Cho and C. Lee, *Nanoscale*, 2018, **10**, 1056–1062.
111 J. Zhang, C. Du, Z. Dai, W. Chen, Y. Zheng, B. Li, Y. Zong, X. Wang, J. Zhu and Q. Yan, *ACS Nano*, 2017, **11**, 10599–10607.
112 M. K. Bahl, *J. Phys. Chem. Solids*, 1975, **36**, 485–491.
113 K. Kim, M. S. Kim, P. R. Cha, S. H. Kang and J. H. Kim, *Chem. Mater.*, 2016, **28**, 1453–1461.
114 S. Ramakrishna, A. Le Viet, M. V. Reddy, R. Jose and B. V. R. Chowdari, *J. Phys. Chem. C*, 2010, **114**, 664–671.
115 R. Romero, J. R. Ramos-Barrado, F. Martin and D. Leinen, *Surf. Interface Anal.*, 2004, **36**, 888–891.
116 M. Aufray, S. Menuel, Y. Fort, J. Eschbach, D. Rouxel and B. Vincent, *J. Nanosci. Nanotechnol.*, 2009, **9**, 4780–4785.
117 D. Gopalakrishnan, A. Lee, N. K. Thangavel and L. M. Reddy Arava, *Sustainable Energy Fuels*, 2018, **2**, 96–102.
118 J. Si, Q. Zheng, H. Chen, C. Lei, Y. Suo, B. Yang, Z. Zhang, Z. Li, L. Lei, Y. Hou and K. Ostrikov, *ACS Appl. Mater. Interfaces*, 2019, **11**, 13205–13213.
119 V. Romano, B. Martín-García, S. Bellani, L. Marasco, J. Kumar Panda, R. Oropesa-Nuñez, L. Najafi, A. E. Del Rio Castillo, M. Prato, E. Mantero, V. Pellegrini, G. D'Angelo and F. Bonaccorso, *ChemPlusChem*, 2019, **84**, 882–892.
120 J. Li, W.-W. Liu, H.-M. Zhou, Z.-Z. Liu, B.-R. Chen and W.-J. Sun, *Rare Met.*, 2018, **37**, 118–122.
121 A. A. Lubimtsev, P. R. C. Kent, B. G. Sumpter and P. Ganesh, *J. Mater. Chem. A*, 2013, **1**, 14951–14956.
122 K. J. Griffith, A. C. Forse, J. M. Griffin and C. P. Grey, *J. Am. Chem. Soc.*, 2016, **138**, 8888–8899.
123 N. Kumagai, Y. Koishikawa, S. Komaba and N. Koshiba, *J. Electrochem. Soc.*, 1999, **146**, 3203–3210.
124 X. Ou, X. Xiong, F. Zheng, C. Yang, Z. Lin, R. Hu, C. Jin, Y. Chen and M. Liu, *J. Power Sources*, 2016, **325**, 410–416.
125 J. Zhang, C. Du, Z. Dai, W. Chen, Y. Zheng, B. Li, Y. Zong, X. Wang, J. Zhu and Q. Yan, *ACS Nano*, 2017, **11**, 10599–10607.
126 B. Stanje, V. Epp, S. Nakhal, M. Lerch and M. Wilkening, *ACS Appl. Mater. Interfaces*, 2015, **7**, 4089–4099.
127 L. Suo, O. Borodin, T. Gao, M. Olguin, J. Ho, X. Fan, C. Luo, C. Wang and K. Xu, *Science*, 2015, **350**, 938–943.
128 Z. Hawash, L. K. Ono, S. R. Raga, M. V. Lee and Y. Qi, *Chem. Mater.*, 2015, **27**, 562–569.
129 Z. Hawash, L. K. Ono and Y. Qi, *Adv. Mater. Interfaces*, 2016, **3**, 1600117.
130 M. Otero, G. Lener, J. Trincavelli, D. Barraco, M. S. Nazzarro, O. Furlong and E. P. M. Leiva, *Appl. Surf. Sci.*, 2016, **383**, 64–70.
131 J. Phillips and J. Tanski, *Int. Mater. Rev.*, 2005, **50**, 265–286.
132 E. Asselin, T. M. Ahmed and A. Alfantazi, *Corros. Sci.*, 2007, **49**, 694–710.
133 R. Subbaraman, D. Tripkovic, D. Strmcnik, K. C. Chang, M. Uchimura, A. P. Paulikas, V. Stamenkovic and N. M. Markovic, *Science*, 2011, **334**, 1256–1260.
134 D. Strmcnik, K. Kodama, D. Van Der Vliet, J. Greeley, V. R. Stamenkovic and N. M. Marković, *Nat. Chem.*, 2009, **1**, 466.

135 D. Strmcnik, M. Escudero-Escribano, K. Kodama, V. R. Stamenkovic, A. Cuesta and N. M. Markovií, *Nat. Chem.*, 2018, **2**, 880.

136 J. Staszak-Jirkovský, C. D. Malliakas, P. P. Lopes, N. Danilovic, S. S. Kota, K.-C. Chang, B. Genorio, D. Strmcnik, V. R. Stamenkovic, M. G. Kanatzidis and N. M. Markovic, *Nat. Mater.*, 2015, **15**, 197.

137 T. Shinagawa, A. T. Garcia-Esparza and K. Takanabe, *Sci. Rep.*, 2015, **5**, 13801.

138 H. Pan, *Sci. Rep.*, 2014, **4**, 5348.

139 J. Lee, S. Kang, K. Yim, K. Y. Kim, H. W. Jang, Y. Kang and S. Han, *J. Phys. Chem. Lett.*, 2018, **9**, 2049–2055.

140 L. Najafi, S. Bellani, R. Oropesa-Nuñez, A. Ansaldo, M. Prato, A. E. Del Rio Castillo and F. Bonaccorso, *Adv. Energy Mater.*, 2018, **8**, 1703212.

141 S. Ma Andersen and E. Skou, *ACS Appl. Mater. Interfaces*, 2014, **61**, 16565–16576.

142 O. J. Curnick, P. M. Mendes and B. G. Pollet, *Electrochem. Commun.*, 2010, **12**, 1017–1020.

143 Z. Zhu, H. Yin, C.-T. He, M. Al-Mamun, P. Liu, L. Jiang, Y. Zhao, Y. Wang, H.-G. Yang, Z. Tang, D. Wang, X.-M. Chen and H. Zhao, *Adv. Mater.*, 2018, **30**, 1801171.

144 R. Subbaraman, D. Tripkovic, K.-C. Chang, D. Strmcnik, A. P. Paulikas, P. Hirunsit, M. Chan, J. Greeley, V. Stamenkovic and N. M. Markovic, *Nat. Mater.*, 2012, **11**, 550.

145 R. Subbaraman, D. Tripkovic, D. Strmcnik, K.-C. Chang, M. Uchimura, A. P. Paulikas, V. Stamenkovic and N. M. Markovic, *Science*, 2011, **334**, 1256–1260.

146 N. Danilovic, R. Subbaraman, D. Strmcnik, K.-C. Chang, A. P. Paulikas, V. R. Stamenkovic and N. M. Markovic, *Angew. Chem.*, 2012, **124**, 12663–12666.

147 F. Bonaccorso, T. Hasan, P. H. Tan, C. Sciascia, G. Privitera, G. Di Marco, P. G. Gucciardi and A. C. Ferrari, *J. Phys. Chem. C*, 2010, **114**, 17267–17285.

148 T. Hasan, P. H. Tan, F. Bonaccorso, A. G. Rozhin, V. Scardaci, W. I. Milne and A. C. Ferrari, *J. Phys. Chem. C*, 2008, **112**, 20227–20232.